\documentclass[prb,aps,amssymb,nofootinbib,twocolumn,showpacs]{revtex4}
\usepackage{amsmath}
\usepackage{amssymb}
\usepackage{amsthm}
\usepackage{amsfonts}
\usepackage{algorithmic}
\usepackage{enumerate}
\usepackage{latexsym}
\usepackage[dvips]{graphicx}

\newcommand{\beq}{\begin{equation}}
\newcommand{\eneq}{\end{equation}}
\newcommand{\beqs}{\begin{equation*}}
\newcommand{\eneqs}{\end{equation*}}
\newcommand{\he}{$^3$He}

\input{epsf}

\begin{document}

\tolerance 10000

\title{Non-Abelian hydrodynamics and the flow of spin in spin-orbit coupled substances.}

\author { B.W.A. Leurs, Z. Nazario, D.I. Santiago and J. Zaanen}

\affiliation{ Instituut Lorentz for Theoretical Physics, Leiden University,
Leiden, The Netherlands}

\begin{abstract}
\begin{center}
Motivated by the heavy ion collision experiments there is much activity in studying the
hydrodynamical properties of non-abelian (quark-gluon) plasma's. A major question is
how to deal with color currents. Although not widely appreciated, quite similar issues arise 
in condensed matter physics in the context of the transport of spins in the presence of spin-orbit
coupling. The key insight  is that the Pauli Hamiltonian governing the leading relativistic corrections in  
condensed matter systems can be rewritten in a language of $SU(2)$ covariant derivatives where the
role of the non-abelian gauge fields is taken by the physical electromagnetic fields: the Pauli system can 
be viewed as Yang-Mills quantum mechanics in a 'fixed frame', and it can be viewed as an 'analogous system' for
non abelian transport in the same spirit as Volovik's  identification of the $He$ superfluids as analogies for  
quantum fields in curved space time. We take a similar perspective as Jackiw and coworkers in their recent 
study of non-abelian hydrodynamics, twisting the interpretation into the 'fixed frame' context, to find out 
what this means for spin transport in  condensed matter systems. We present an extension of Jackiw's scheme: 
non-abelian hydrodynamical currents can be factored in  a 'non-coherent' classical
part, and a coherent part requiring macroscopic non-abelian quantum entanglement.  Hereby it becomes
particularly manifest that non-abelian fluid flow is a much richer affair than familiar hydrodynamics, and this
permits us to classify the various spin transport phenomena in condensed matter physics in an unifying framework.
 The "particle based hydrodynamics" of Jackiw {\em et al.} is recognized as the high temperature
spin transport associated with semiconductor spintronics. In this context the absence of faithful hydrodynamics 
is
well known, but in our formulation it is directly associated with the fact that the covariant conservation
of non-abelian currents turns into a disastrous non-conservation of the incoherent spin currents of the
high temperature limit. We analyze the quantum-mechanical 
single particle currents of relevance to mesoscopic transport with as highlight the Ahronov-Casher effect,
where we demonstrate that the intricacies of the non-abelian transport render this effect to be much more
fragile than its abelian analogue, the Ahronov-Bohm effect. We subsequently focus on spin flows protected
by order parameters. At present there is much interest in multiferroics where non-collinear magnetic order 
triggers macroscopic electric polarization via the spin-orbit coupling. We identify this to be a peculiarity of
coherent non-abelian hydrodynamics: although there is no net particle transport, the spin entanglement is
transported in these magnets and the coherent spin 'super' current   in turn translates into electric fields
with the bonus that due to the requirement of single valuedness of the magnetic order parameter a true
hydrodynamics is restored. Finally,'fixed-frame' coherent non-abelian transport comes to its full glory in 
spin-orbit 
coupled 'spin superfluids',  and we demonstrate  a new effect: the trapping of electrical line charge being a 
fixed 
frame, non-abelian analogue of the familiar magnetic flux trapping by normal superconductors. The only known  
physical examples of such spin superfluids are the $^3$He A- and B phase where unfortunately the spin-orbit 
coupling is so weak that it appears impossible to observe these effects. 
\end{center}
\end{abstract}

\date{\today}


\maketitle

\section{Introduction}

It is a remarkable development that in various branches of physics there is a revival going
on of the long standing problem of how non-Abelian entities are transported over
macroscopic
distances. An important stage is condensed matter physics. A first major development is
spintronics, the pursuit to use the electron spin instead of its charge for switching
purposes\cite{murakami03,murakami04, culcer04, sinova04, mishchenko03, mishchenko04},
with a main focus on transport in conventional semiconductors. Spin-orbit coupling is
needed to create and manipulate these spin currents , and it has become increasingly clear that
transport phenomena are possible that are quite different from straightforward electrical
transport.
A typical example is the spin-Hall effect\cite{murakami03, murakami04, sinova04},
defined
through the macroscopic transport equation,
    \begin{equation}
      j^a_i = \sigma_{SH} \epsilon_{ial} E_l
      \label{spinhall}
    \end{equation}
where $\epsilon_{ial}$ is the 3-dimensional Levi-Civita tensor and $E_l$ is the electrical
field. The specialty is that since both $j^a_i$ and $E_l$ are even under time reversal, the
transport coefficient $\sigma_{SH}$ is also even under time reversal, indicating that this
corresponds with a dissipationless transport phenomenon.  An older
development is the mesoscopic spin-transport analogue of the Aharonov-Bohm effect, called the
Aharonov-Casher effect\cite{aharonovcasher}: upon transversing a loop containing an electrically charged
wire the spin conductance will show oscillations with a period set by the strength of the
spin-orbit coupling and the enclosed electrical line-charge.

A rather independent development in condensed matter physics is the recent focus on the
multiferroics. This refers to substances that show simultaneous ferroelectric- and
ferromagnetic order at low temperatures, and these two different types of order do
rather strongly depend on each other. It became clear recently that at least in an
important subclass of these systems  one can explain the phenomenon in a language
invoking dissipationless spin transport\cite{nagaosabalatskii,mostovoy}: one needs a
magnetic order characterized by spirals such that 'automatically' spin currents are
flowing, that in turn via spin-orbit coupling induce electrical fields responsible for the
ferroelectricity.

The final condensed matter example is one that was lying dormant over the last years: the
superfluids realized in \he . A way to conceptualize the intricate order parameters of the A-
and B-phase\cite{volovikexo,leggetthe} is to view these as non-Abelian ('spin-like')
superfluids. The intricacies of the topological defects in these phases is of course very well
known, but matters get even more interesting when considering the effects on the superflow of
macroscopic electrical fields, mediated by the very small but finite spin-orbit coupling. This
subject has been barely studied: there is just one paper by Mineev and Volovik\cite{minvol}
addressing these matters systematically.

A very different pursuit is the investigation of the quark-gluon plasma's presumably generated
at the Brookhaven heavy-ion collider. This might surprise the reader: what is the relationship
between the flow of spin in the presence of spin-orbit coupling in the cold condensed matter
systems and this high temperature QCD affair? There is actually a very deep connection that
was already realized quite some time ago. Goldhaber\cite{goldhaber} and later Froehlich {\em
et al}\cite{frohlich}, Balatskii and Altshuler\cite{balatskiialtshuler} and others realized
that in the presence of spin-orbit coupling spin is subjected to a parallel transport
principle that is quite similar to the parallel transport of matter fields in Yang-Mills
non-Abelian gauge theory, underlying for instance QCD. This follows from a simple rewriting of
the Pauli-equation, the Schroedinger equation taking into account the leading relativistic
corrections: the spin-fields are just subjected to covariant derivatives of the Yang-Mills
kind, see Eq.'s (\ref{Di}),(\ref{D0}). However, the difference is that the 'gauge' fields
appearing in these covariant derivatives are actually physical fields. These are just
proportional to the electrical- and magnetic fields. Surely, this renders the problem of spin
transport in condensed matter systems to be dynamically very different from the fundamental
Yang-Mills theory of the standard model. However, the parallel transport structure has a 'life of
its own': it implies certain generalities that are even independent of the 'gauge' field being
real gauge or physical.

For all the examples we alluded to in the above, one is dealing with macroscopic
numbers of particles that are collectively transporting non-Abelian quantum numbers over
macroscopic distances and times. In the Abelian realms of electrical charge or mass a
universal description of this transport is available in the form of hydrodynamics, be it the
hydrodynamics of water, the magneto-hydrodynamics of charged plasma's, or the
quantum-hydrodynamics of superfluids and superconductors. Henceforth, to get
anywhere in terms of a systematic description one would like to know how to think in a
hydrodynamical fashion about the macroscopic flow of non-Abelian entities, including
spin.

In the condensed matter context one finds pragmatic, case to case approaches that are
not necessarily wrong, but are less revealing regarding the underlying 'universal'
structure: in spintronics one solves Boltzmann transport equations, limited to dilute and
weakly interacting systems. In the quark-gluon plasma's one find a similar attitude,
augmented by RPA-type considerations to deal with the dynamics of the gauge fields. In
the multiferroics one rests on a rather complete understanding of the order parameter
structure.

The question remains: what is non-Abelian hydrodynamics? To the best of our knowledge this
issue is only addressed on the fundamental level by Jackiw and coworkers \cite{jackiw1, jackiw2}
 and their work forms
a main inspiration for this review. The unsettling answer seems to be: {\em non-Abelian
hydrodynamics in the conventional sense of describing the collective flow of quantum numbers
in the classical liquid does not even exist!} The impossibility to define 'soft'
hydrodynamical  degrees of freedom is rooted in the non-Abelian parallel transport structure
per se and is therefore shared by high temperature QCD and spintronics.

The root of the trouble is that non-Abelian currents do not obey a continuity equation but are
instead only {\em covariantly conserved}, as we will explain in detail in section
\ref{spincurrcovcons}. It is well known that covariant conservation laws do not lead to global
conservation laws, and the lack of globally conserved quantities makes it impossible to deal
with matters in terms of a universal hydrodynamical description. This appears to be a most
serious problem for the description of the 'non-Abelian fire balls' created in Brookhaven. In
the spintronics context it is well known under the denominator of 'spin relaxation': when a
spin current is created, it will plainly disappear after some characteristic spin relaxation
determined mostly by the characteristic spin-orbit coupling strength of the material.

In this review we will approach the subject of spin transport in the presence of spin-orbit
coupling from the perspective of the non-Abelian parallel transport principle. At least to
our perception, this makes it possible to address matters in a rather unifying, systematical
way. It is not a-priori  clear how the various spin transport phenomena identified in
condensed matter relate to each other and we hope to convince the reader that they are
different sides of the same non-Abelian hydrodynamical coin. Except for the inspiration
we have found in the papers by Jackiw and coworkers \cite{jackiw1, jackiw2} we will largely 
ignore the subject
of the fundamental non-Abelian plasma, although we do hope that the 'analogous systems'
we identify in the condensed matter system might form a source of inspiration for those
working on the fundamental side.

Besides bringing some order to the subject, in the course of the development we found
quite a number of new and original results that are consequential for the general, unified
understanding. We will start out on the pedestrian level of quantum-mechanics (section III),
discussing in detail how the probability densities of non-Abelian quantum numbers are
transported by isolated quantum particles and how this relates to spin-orbit coupling
(Section IV). We will derive here equations that are governing the mesoscopics, like the
Aharonov-Casher (AC) effect, in a completely general form. A main conclusion will
be that already on this level the troubles with the macroscopic hydrodynamics are
shimmering through:  the AC effect is more fragile than the Abelian Aharonov-Bohm
effect, in the sense that the experimentalists have to be much more careful in designing
their machines in order to find the AC signal.

In the short section V we revisit the non-Abelian covariant conservation laws, introducing a
parametrization that we perceive as very useful: different from the Abelian case, non-Abelian
currents can be viewed as being composed of both a coherent, 'spin' entangled part and a
factorisable incoherent part. This difference is at the core of our classification of
non-Abelian fluids. The non-coherent current is responsible for the transport in the high
temperature liquid. The coherent current is responsible for the multiferroic effects,
the Meissner 'diamagnetic' screening currents in the fundamental non-Abelian Higgs
phase, but also for the non-Abelian supercurrents in true spin-superfluids like the \he\ A-
and B phase.

The next step is to deduce the macroscopic hydrodynamics from the microscopic constituent
equations and here we follow Jackiw {\it et. al.}\cite{jackiw1, jackiw2} closely. Their 'particle based' non-Abelian
hydrodynamics is just associated with the classical hydrodynamics of the high
temperature spin-fluid and here the lack of hydrodynamical description hits full force: we
hope that the high energy physicists find our simple 'spintronics' examples illuminating
(Section VI).

After a short technical section devoted to the workings of electrodynamics in the SO problem (section VII),
we turn to the 'super' spin currents of the multiferroics (Section VIII).
As we will show, these are rooted in the coherent non-Abelian currents and this
renders it to be quite similar but subtly different from the 'true' supercurrents of the spin
 superfluid:
 it turns
out that in contrast to the latter they can create electrical charge!  This is also a most
elementary context to introduce a notion that we perceive as the most important feature of
non-Abelian fluid theory. In Abelian hydrodynamics it is well understood when the superfluid
order sets in, its rigidity does change the hydrodynamics: it renders the hydrodynamics of the
superfluid to be irrotational having the twofold effect that the circulation in the superfluid
can only occur in the form of massive, quantized vorticity while at low energy the superfluid
is irrotational so that it behaves like a dissipationless ideal Euler liquid. In the
non-Abelian fluid the impact of the order parameter is more dramatic: its rigidity removes the
multivaluedness associated with the covariant derivatives and hydrodynamics is restored!

This bring us to our last subject where we have most original results to offer: the
hydrodynamics of spin-orbit coupled spin-superfluids (Section IX). These are the 'fixed
frame' analogs of the non-Abelian Higgs phase and we perceive them as the most beautiful
physical species one encounters in the non-Abelian fluid context. Unfortunately, they do not
seem to be prolific in nature. The \he\ superfluids belong to this category but it is an
unfortunate circumstance that the spin-orbit coupling is so weak that one encounters
insurmountable difficulties in the experimental study  of its effects. Still we will use them
as an exercise ground to demonstrate how one should deal with more complicated non-Abelian
structures (Section XI), and we will also address the issue of where to look for other
spin-superfluids in the concluding section (Section XII).

To raise the appetite of the reader let us start out presenting some wizardry that should be
possible to realize in a laboratory when a spin-superfluid would be discovered with a
sizable spin-orbit coupling: how the elusive spin-superfluid manages to trap electrical line
charge (section II), to be explained in detail in Section X.

\section{The Appetizer: trapping quantized electricity.}

\begin{figure}[ht!]
      \centering
      \rotatebox{0}{
    \resizebox{5.3cm}{!}{%
      \includegraphics*{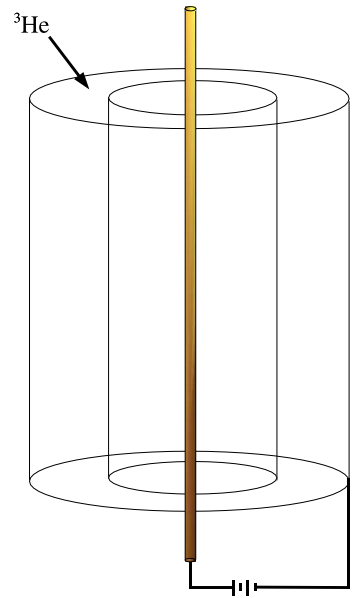}}}
      \caption{ A superfluid $^3$He container acts as a capacitor capable
      of trapping a quantized electrical line charge density via the electric
      field generated by persistent spin Hall currents. This is te analog
      of magnetic flux trapping in superconductors by persistent
      charge supercurrents.} \label{experiment}
\end{figure}

Imagine a cylindrical vessel, made out of plastic while its walls are coated with a thin layer
of gold.  Through the center this vessel a gold wire is threaded and care is taken that it is
not in contact with the gold on the walls. Fill this container to the brim with a putative liquid that 
can become a spin superfluid (liquid \he\ would work if it did not contain a dipolar interaction 
that voids the physics ) in
its normal state and apply now a large bias to the wire keeping the walls grounded, see Fig.
\ref{experiment}. Since it is a capacitor, the wire will charge up relative to the walls. Take
care that the line charge density on the wire is pretty close to a formidable $2.6 \times
10^{-5}$ Coulomb per meter in the case that this fluid would be like \he\ .

Having this accomplished, cool the liquid through its spin superfluid phase transition temperature
$T_c$.  Remove now the voltage and hold  the end of the wire close to the vessel's wall.
Given that the charge on the wire is huge, one anticipates a disastrous decharging spark
but .... nothing happens!

It is now time to switch off the dilution fridge. Upon monitoring the rising temperature, right
at  $T_c$ where the spin superfluid turns normal a spark jumps from the wire to the vessel,
grilling the machinery into a pile of black rubble.

This is actually a joke. In Section X we will present the theoretical proof that
this experiment can actually be done. There is a caveat, however. The only substance that has
been identified, capable of doing this trick is helium III were it not for the dipolar interaction preventing it 
being the desired spin superfluid. But even if we were God and we could turn the
dipolar locking to zero making Helium III into the right spin superfluid, there would still be trouble.
In order to
prevent bad things to happen {\em one needs a vessel with a cross sectional area that is
roughly equal to the area of Alaska}. Given that there is only some 170 kg of helium on our
planet, it occurs that this experiment cannot be practically accomplished.

What is going on here? This effect is  analogous to magnetic flux trapping by
superconducting rings. One starts out there with the ring in the normal state, in the
presence of an external magnetic field. One cycles the ring below the transition
temperature, and after switching off the external magnetic field a quantized magnetic
flux is trapped by the ring. Upon cycling back to the normal state this flux is expelled.
Read for the magnetic flux the electrical line charge, and for the electrical
superconductor the spin-superfluid and the analogy is clear.

This reveals that in both cases a similar parallel transport principle is at work. It is surely not so
that this can be understood by simple electro-magnetic duality: the analogy is imprecise
because of the fact that the physical field enters in the spin-superfluid problem via the
spin-orbit coupling in the same way the vector potential enters in superconductivity. This has the
ramification that the electrical monopole density takes the role of the magnetic flux, where
the former takes the role of physical incarnation of the pure gauge Dirac string associated
with the latter.

The readers familiar with the Aharonov-Casher effect should hear a bell ringing\cite{balatskiialtshuler}. This can
indeed be considered as just the 'rigid' version of the AC effect, in the same way that
flux trapping is the rigid counterpart of the mesoscopic Aharonov-Bohm effect. On the
single particle level, the external electromagnetic fields prescribe the behavior of the
particles, while in the ordered state the order parameter has the power to impose its
will on the electromagnetic fields.

This electrical line-charge trapping effect summarizes neatly the deep but incomplete
relations between real gauge theory and the working of spin-orbit coupling. It will
be explained in great detail in sections IX and X, but before we get there we first
have to cross some terrain.

\section{Quantum Mechanics of Spin-Orbit Coupled Systems}

To address the transport of spin in the presence of spin-orbit (SO) coupling we will follow a
strategy well known from conventional quantum mechanical transport theory. We will first
analyze the single particle quantum-mechanical probability currents and densities. The
starting point is the Pauli equation, the generalization of the Schr\"odinger equation
containing the leading relativistic corrections as derived by expanding the Dirac equation
using the inverse electron rest mass as expansion parameter. We will first review the
discovery by Volovik and Mineev \cite{minvol}, Balatskii and Altshuler
\cite{balatskiialtshuler} and Froehlich and others \cite{frohlich} of the
non-Abelian parallel transport structure hidden in this equation, to subsequently analyze in
some detail the equations governing the spin-probability currents. In fact, this is closely
related to the transport  of color currents in real Yang-Mills theory: the fact that in the SO
problem the 'gauge fields' are physical fields is of secondary importance since  the most
pressing issues regarding non-Abelian transport theory hang together with parallel transport.
For these purposes, the spin-orbit 'fixed-frame' incarnation has roughly the status as a
representative gauge fix. In fact, the development in this section has a substantial overlap
with the work of Jackiw and co-workers dedicated to the development of a description of
non-Abelian fluid dynamics \cite{jackiw1,jackiw2}. We perceive the application to the specific
context of SO coupled spin fluid dynamics as clarifying and demystifying in several regards.
We will identify their 'particle based' fluid dynamics with the high temperature, classical
spin fluid where the lack of true hydrodynamics is well established, also experimentally.
Their 'field based' hydrodynamics can be directly associated with the coherent superflows
associated with the SO coupled spin superfluids where at least in equilibrium a sense of a
protected hydrodynamical sector  is restored.

The development in this section have a direct relevance to mesoscopic transport
phenomena (like the Aharonov-Casher effects\cite{aharonovcasher,balatskiialtshuler}, but
here our primairy aim is to set up the system of microscopic, constituent equations
to be used in the subsequent sections to derive the various macroscopic fluid theories.
The starting point is the well known Pauli-equation describing mildly relativistic particles.
This can be written in the form of a Lagrangian density in terms of spinors $\psi$,
      \begin{align}
      \begin{aligned}
    \mathcal L &= i\hbar\psi^\dagger(\partial_0\psi) -
    qB^a\psi^\dagger\frac{\tau^a}{2}\psi +
    \frac{\hbar^2}{2m}\psi^\dagger\left( \nabla
    -\frac{ie}{\hbar}\vec A \right)^2\psi \\
    &-eA_0\psi^\dagger\psi + \frac{iq}{2m}\epsilon_{ial}E_l
    \left\{ (\partial_i\psi^\dagger)\frac{\tau^a}{2}\psi -
    \psi^\dagger\frac{\tau^a}{2}(\partial_i\psi) \right\} \\
    \label{lag}&+ \frac{1}{8\pi}\left( E^2-B^2 \right)
      \end{aligned}
    \end{align}
where 
    \beq
      \vec E = - \nabla A_0 - \partial_0 \vec A \, , \qquad \vec B =
      \nabla\times\vec A
    \eneq
$A_\mu$ are the usual $U(1)$ gauge fields associated with  the electromagnetic fields
$\vec E$ and $\vec B$. The relativistic corrections are present in the terms containing
the quantity  $q$, proportional to the Bohr magneton, and the time-like first term $\propto
B$ is the usual Zeeman term while the space-like terms $\propto E$ corresponds with
spin-orbital coupling.

The recognition that this has much to do with a non-Abelian parallel transport structure, due
to Mineev and Volovik \cite{minvol}, Goldhaber \cite{goldhaber}
 and Froehlich et al. \cite{frohlich} is in fact very simple. Just redefine the
magnetic- and electric field strengths as follows,
    \beq
      \label{map}
      A_0^a = B^a \qquad \qquad A_i^a = \epsilon_{ial}E_l \, ,
    \eneq
Define covariant derivatives as usual,
    \begin{align}
      D_i&=\partial_i - i \frac{q}{\hbar} A_i^a \frac{\tau^a}{2} -i
      \frac{e}{\hbar} A_i \label{Di} \\
      D_0&=\partial_0 + i \frac{q}{\hbar} A_0^a \frac{\tau^a}{2} + i
      \frac{e}{\hbar} A_0  \label{D0} \, .
    \end{align}
and it follows that the Pauli equation in Lagrangian form becomes,
    \begin{align}
      \begin{aligned}
    \nonumber \mathcal L &= i\hbar \psi^\dagger D_0\psi +
    \psi^\dagger\frac{\hbar^2}{2m}\vec D^2\psi \\
    &+ \frac{1}{2m}\psi^\dagger\left(2eq\frac{\tau^a}{2}\vec
    A\cdot\vec A^a + \frac{q^2}{4}\vec A^a\cdot\vec A^a\right)\psi
    \\
    \label{covlag}&+ \frac{1}{8\pi}\left(E^2-B^2\right) \, .
      \end{aligned}
    \end{align}
Henceforth, the derivatives are replaced by the covariant derivatives of a $U(1) \times
SU(2)$ gauge theory, where the $SU(2)$ part takes care of the transport of spin. Surely,
the second and especially the third term violate the $SU(2)$ gauge invariance for the
obvious reason that the non-Abelian 'gauge fields' $A^a_{\mu}$ are just proportional to
the electromagnetic $\vec{E}$ and $\vec{B}$ fields. Notice that the second term just amounts
to a small correction to the electromagnetic part (third term).  The standard picture of how
spins are precessing due to the spin-orbit coupling to external
electrical- and magnetic fields, pending the way they are moving through space can
actually be taken as a literal cartoon of the parallel transport of non-Abelian charge in
some fixed gauge potential!

To be more precise, the SO problem does actually correspond with a particular gauge fix
in the full $SU(2)$ gauge theory. The electromagnetic fields have to obey  the Maxwell
equation,
    \beq
      \nabla\times\vec E + \frac{\partial \vec B}{\partial t} = 0
    \eneq
and this in turn implies
    \beq
      \partial^\mu A_\mu^a = 0 \; .
    \eneq
Therefore,  the SO problem is 'representative' for the $SU(2)$ gauge theory in
the Lorentz gauge and we do not have the choice of going to another gauge
as the non-Abelian fields are expressed in terms of real electric and magnetic fields.
This is a first new result.

By varying the Lagrangian with respect to $\psi^\dagger$ we obtain the Pauli equation in
its standard Hamiltonian form,
    \beq
      \label{paulilike}
      i\hbar D_0\psi = -\frac{\hbar^2}{2m} D_i^2\psi -
      \frac{1}{2m}\left(2eq\frac{\tau^a}{2}\vec A\cdot\vec A^a +
      \frac{q^2}{4}\vec A^a\cdot\vec A^a\right)\psi
    \eneq
where we leave the electromagnetic part implicit, anticipating that we will be interested to
study the behavior of the quantum mechanical particles in fixed background
electromagnetic field configurations. The wave function $\psi$ can be written in the form,
    \beq \label{spinor}
      \psi = \sqrt\rho \; e^{(i\theta + i\varphi^a\tau^a/2)} \chi
    \eneq
with the probability density $\rho$, while $\theta$ is the usual Abelian phase associated with
the electromagnetic gauge fields. As imposed by the covariant derivatives, the $SU(2)$ phase
structure can be parametrised by the three non-Abelian phases $\varphi^a$, with the Pauli
matrices $\tau^a$ acting on a reference spinor $\chi$. Hence, with regard to the wavefunction
there is no difference whatever between the Pauli-problem and genuine Yang-Mills quantum
mechanics: this is all ruled by parallel transport.

Let us now investigate in further detail how the Pauli equation transports spin-probability.
This is in close contact with work in high-energy physics and we develop the theory along similar
lines as Jackiw{\em et al.}\cite{jackiw2}. We introduce, however, a condensed matter inspired
parametrization that we perceive as instrumental towards laying bare the elegant meaning of
the physics behind the equations. 

A key ingredient of our parametrization is the introduction of a non-Abelian phase
velocity, an object  occupying the adjoint together with the vector potentials.The
equations in the remainder will involve time and space derivatives of $\theta$, $\rho$ and
of the spin rotation operators
    \beq
      e^{i\varphi^a\tau^a/2} \; .
    \eneq
Let us introduce the operator  $S^a$ as the non-Abelian charge at time
$t$ and at position $\vec r$, as defined by the appropriate $SU(2)$ rotation
    \beq \label{Sa}
      S^a \equiv  e^{-i\varphi^a\tau^a/2} \; \frac{\tau^a}{2} \;
      e^{i\varphi^a\tau^a/2} \, .
    \eneq
The temporal and spatial dependence arises through the non-Abelian phases
$\varphi^a(t,\vec r)$. The non-Abelian charges are, of course, $SU(2)$ spin $1/2$
operators:
    \beq
      S^a S^b = \frac{\delta^{ab}}{4} + \frac{i}{2}\epsilon^{abc}S^c
    \eneq
 It is illuminating to parametrize the derivatives of the spin rotation operators employing
non-Abelian velocities $\vec{u}^a$ defined by,
    \begin{align}
      \begin{aligned} \label{nonabveli}
    &\frac{im}{\hbar}\vec u^a S^a \equiv
    e^{-i\varphi^a\tau^a/2}(\nabla e^{i\varphi^a\tau^a/2}) \qquad
    \text{ or} \\
    &\vec u^a = -2i\frac{\hbar}{m} \text{Tr}
    \left\{e^{-i\varphi^a\tau^a/2}(\nabla e^{i\varphi^a\tau^a/2})
    S^a \right\} \, ,
      \end{aligned}
    \end{align}
which are  just the analogs of the usual Abelian phase velocity
    \beq
      \vec u \equiv \frac{\hbar}{m}\nabla\theta = -i\frac{\hbar}{m}
      e^{-i\theta} \nabla e^{i\theta}\, .
    \eneq
These non-Abelian phase velocities represent the scale parameters for the propagation of spin
probability in non-Abelian quantum mechanics, or either for the hydrodynamical flow of
spin-superfluid.

In addition we need the zeroth component of the velocity
    \begin{align}
      \begin{aligned}\label{nonabvel0}
    i u^a_0 S^a \equiv e^{-i\varphi^a\tau^a/2}(\partial_0
    e^{i\varphi^a\tau^a/2}) \qquad \text{ or} \\
    u^a_0 = -2 i \text{Tr}
    \left\{e^{-i\varphi^a\tau^a/2}(\partial_0
    e^{i\varphi^a\tau^a/2}) S^a \right\}
      \end{aligned}
    \end{align}
being the time rate of change of the non-Abelian phase,  amounting to a precise analog of the
time derivative of the Abelian phase representing matter-density fluctuation,
    \beq
      u_0 \equiv \partial_0 \theta = -i\frac{\hbar}{m} e^{-i\theta}
      \partial_0 e^{i\theta}\; .
    \eneq
 It is straightforward to show that  the definitions of the spin operators $S^a$, Eq.(\ref{Sa}) and
the non-Abelian velocities $u_\mu^a$, Eq.'s(\ref{nonabveli}, \ref{nonabvel0}), imply in
combination,
    \beq
      \partial_0 S^a= - \epsilon^{abc}u^b_0 S^c \quad \nabla S^a= -
      \frac{m}{\hbar}\epsilon^{abc}\vec u^b S^c
    \eneq

It is easily checked that the definition of the phase velocity Eq. (\ref{nonabveli}) implies the
following identity,
    \beq
      \nabla \times \vec u^a + \frac{m}{2\hbar}\epsilon_{abc}\vec u^b
      \times \vec u^c=0 \; ,
    \eneq
having as Abelian analogue,
    \beq
      \nabla \times \vec u =0 \; .
    \eneq
as the latter controls vorticity, the former is in charge of the topology in the non-Abelian
'probability fluid'. It, however, acquires a truly quantum-hydrodynamical status in the rigid
superfluid where it becomes an equation of algebraic topology. This equation is
well known, both in gauge theory and in the theory of the \he\ superfluids where it is known
as the Mermin-Ho equation\cite{merminho}.

\section{Spin transport in the mesoscopic regime}

Having defined the right variable, we can now go ahead with the quantum mechanics, finding
transparent equations for the non-Abelian probability transport. Given that this is about
straight quantum mechanics, what follows does bare relevance to coherent spin transport
phenomena in the mesoscopic regime. We will actually derive some interesting results that
reveal subtle caveats regarding mesoscopic spin transport. The punchline is that the
Aharonov-Casher effect and related phenomena are intrinsically fragile, requiring much more
fine tuning in the experimental machinery than in the Abelian (Ahronov-Bohm) case.

Recall the spinor definition Eq.(\ref{spinor}); together with the definitions of the phase
velocity, it follows from the vanishing of the imaginary part of the Pauli equation
that,
    \beq
      \label{cons0}
      \partial_0 \rho + \vec \nabla \cdot \left[\rho  \left( \vec u
    -\frac{e}{m} \vec A + \vec u^a S^a - \frac{q}{m}\vec A^a S^a
    \right)  \right]=0
    \eneq
and this is nothing else than the non-Abelian continuity equation, imposing that probability
is covariantly conserved. For non-Abelian parallel transport this is a weaker condition
than for the simple Abelian case where the continuity equation implies a global
conservation of mass, being in turn the condition for hydrodynamical degrees of freedom
in the fluid context. Although locally conserved, the non-Abelian charge is not globally
conserved and this is the deep reason for the difficulties with associating a universal
hydrodynamics to the non-Abelian fluids.  The fluid dynamics will borrow this motive
directly from quantum mechanics where its meaning is straightforwardly isolated.

Taking the trace over the non-Abelian labels in Eq. (\ref{cons0}) results in the usual
continuity equation for Abelian probability, in the spintronics context associated with the
conservation of electrical charge,
    \beq
      \label{consn}
      \partial_0\rho + \nabla\cdot\left[\rho \left(\vec u -\frac{e}{m}
      \vec A \right)\right] = 0 \, ,
    \eneq
 where one recognizes the standard (Abelian) probability current,
    \beq
      \label{curra}
      \vec J = \rho\left(\vec u -\frac{e}{m} \vec A \right)=
      \frac{\hbar}{m}\rho\left(\nabla\theta - \frac{e}{\hbar}\vec
      A\right) \, .
    \eneq
 From  Abelian continuity and the full non-Abelian law Eq. (\ref{cons0}) it is directly
seen that the non-Abelian velocities and vector potentials have to satisfy the following
equations,
    \beq
    \nabla\cdot \left[ \rho\left(\vec u^a - \frac{q}{m}\vec
      A^a\right) \right]  = \frac{q}{\hbar}\rho\epsilon^{abc} \vec
      u^b\cdot\vec A^c
      \label{messphall}
  \eneq
and we recognize a divergence -- the quantity inside the bracket is a conserved,
current-like quantity. Notice that in this non-relativistic theory this equation contains
only space like derivatives: it is a static constraint equation stating that the non-Abelian
probability density should not change in time.
 The above is generally valid but it is instructive to now interpret this result
in the Pauli-equation context. Using Eq.(\ref{map}) for the non Abelian vector potentials,
 Eq (\ref{messphall}) becomes,
   \begin{equation} \label{consna}
\partial_i \left[ \rho\left( u^a_i - \frac{q}{m} \epsilon_{ail} E_l
      \right) \right]  = - \frac{q}{\hbar}\rho  \left(u^b_a E_b - u^b_b E_a \right)
   \end{equation}
As a prelude to what is coming, we find that  this actually amounts to a statement about
spin Hall probability currents. When the quantity on the r.h.s. would be zero,
$j^a_i = \rho u^a_i =  \frac{\rho q}{m} \epsilon_{ail} E_l + \nabla \times \vec \lambda$,
the spin Hall equation modulo an arbitrary curl and thus the spin Hall relation exhibits
a ``gauge invariance''.

Let us complete this description of non-Abelian quantum mechanics by inspecting the
real part of the Pauli equation in charge of the time evolution of the phase,
\begin{align}
\begin{aligned}
    &  \partial_0\theta   -e A_0 + u_0^aS^a
      - q A_0^a S^a\\
      & = -\frac{1}{\hbar} \left( \frac{m}{2} \left[ \vec u
      -\frac{e}{m} \vec A + \vec u^a S^a - \frac{q}{m} \vec A^a
      S^a \right]^2 \right. \\
    &   +
    \left. \frac{1}{2m}\left[2eqS^a\vec A\cdot\vec A^a
      + \frac{q^2}{4}\vec A^a\cdot\vec A^a\right] \right) \\
 & +
      \frac{\hbar}{4m} \left[ \frac{\nabla^2\rho}{\rho} -
      \frac{(\nabla\rho)^2}{2\rho^2} \right]  . \label{joe}
    \end{aligned}
    \end{align}
 Tracing out the non-Abelian sector we obtain the usual equation for the time rate of
change of the Abelian phase, augmented by two $SU(2)$ singlet terms on the r.h.s.,
    \begin{align}
      \begin{aligned}
    \label{joea}
    & \partial_0\theta -e A_0\; = \frac{\hbar}{4m} \left[
    \frac{\nabla^2\rho}{\rho} - \frac{(\nabla\rho)^2}{2\rho^2}
    \right]\\
    & -\frac{1}{\hbar} \left( \frac{m}{2} \left[ \left(\vec u
    -\frac{e}{m} \vec A \right)^2 + \frac{1}{4}\vec u^a \cdot \vec
    u^a - \frac{q}{2m} \vec u^a \cdot \vec A^a \right] \right) \, .
      \end{aligned}
    \end{align}
Multiplying this equation by $S^b$ and tracing the non-Abelian labels we find,
    \beq
      \label{joena}
      u_0^a - q A_0^a = -\frac{m}{\hbar} \left( \vec u -\frac{e}{m}
      \vec A \right)\cdot \left(\vec u^a - \frac{q}{m} \vec A^a\right)
    \eneq
It is again instructive to consider the spin-orbit coupling interpretation,
  \beq
      \label{joenb}
      u_0^a  = qB_a  - \frac{m}{\hbar}  \left( u_i - \frac{e}{m}
      \vec A_i \right) \cdot \left(u_i^a - \frac{q}{m} \epsilon_{ial} E_l \right)
    \eneq
ignoring the spin orbit coupling this just amounts to Zeeman coupling. The second term on
the right hand side is expressing that spin orbit coupling can generate uniform
magnetization, but this requires both matter current (first term) and a {\em violation}
of the spin-Hall equation! As we have just seen such violations, if present, {\it necessarily}
take the form of a curl.

To appreciate further what these equations mean, let us consider an experiment of the
Aharonov-Casher\cite{aharonovcasher} kind. The experiment consists of an electrical 
wire oriented, say,
along the z-axis that is charged, and is therefore producing an electrical field $E_r$ in the
radial direction in the xy plane. This wire is surrounded by a loop containing mobile
spin-carrying but electrically neutral particles (like neutrons  or atoms ).
Consider now the spins of the particles to be polarized along the z-direction and it is
straightforward to demonstrate that the particles accumulate a holonomy $\sim E_r$. It is
easily seen that this corresponds with a special case in the above formalism. By specializing
to spins lying along the z-axis, only one component $\vec{u}^z, u^z_0$ of the non-Abelian
phase velocity  $\vec{u}^a, u^a_0$ has to be considered, and this reduces the problem to a
$U(1)$ parallel transport structure; this reduction is rather implicit in the standard
treatment.

Parametrise the current loop in terms of a radial ($r$) and azimuthal ($\phi$) direction.
Insisting that the electrical field is entirely along $r$, while the spins are oriented along
$z$ and the current flows in the $\phi$ direction so that only $u_{\phi}^z \neq 0$ , Eq.
(\ref{consna}) reduces to $\partial_{\phi}\left(  \rho ( u^z_\phi - (q/m) E_r ) \right)  = 0$.
$J^z_{\phi} = \rho u^z_\phi$ corresponds with a spin probability current, and it follows that
$J^z_{\phi}= ( q \rho/m) E_r + f(r,z)$ with $f$ an arbitrary function of the vertical and
radial coordinates: this is just the quantum-mechanical incarnation of the spin-Hall transport
equation Eq. (\ref{spinhall})! For a very long wire in which all vertical coordinates are
equivalent, the cylindrical symmetry imposes $z$ independence, and since we are at fixed
radius, $f$ is a constant. In the case where the constant can dropped we have $u^z_{\phi} =
\partial_{\phi} \theta^z = (q/m) E_r$ the phase accumulated by the particle by moving around
the loop equals $\Delta \theta^z = \oint d\phi u^z_{\phi} = L (q/m) E_r$: this is just the
Aharonov-Casher phase. There is the possibility that the Aharonov-Casher effect might not
occur if physical conditions make the constant $f$ nonzero.

Inspecting the 'magnetization' equation Eq. (\ref{joenb}), assuming there is no magnetic
field while the particle carries no electrical charge, $u^a_0 = - ( m/ \hbar) \vec{u}
\cdot (\vec{u}^a - (q/m) \epsilon_{ial} E_l) = 0$, given the conditions of the ideal
Aharonov-Casher experiment. Henceforth, the spin currents in the AC experiment do
not give rise to magnetization.

The standard AC effect appears to be an outcome of a rather special, in fact fine tuned
experimental geometry, hiding the intricacies of the full non-Abelian situation expressed by
our equations Eq. (\ref{consna},\ref{joenb}). As an example, let us consider the simple
situation that, as before, the spins are polarized along the z-direction while the current
flows along $\phi$ such that only $u^z_{\phi}$ is non zero. However, we assume now a stray
electrical field along the z-direction, and it follows from Eq. (\ref{consna}),
  \beq
   \partial_{\phi} \left(  \rho ( u^z_\phi - \frac{q}{m} E_r ) \right)  = - \frac {q}{\hbar}
u^z_{\phi} E_z
  \eneq
We thus see that if the field is not exactly radial, the nonradial parts will provide corrections to
the spin Hall relation and more importantly will invalidate the Aharonov-Casher effect!
This stray electrical field in the z-direction has an even simpler implication for the
magnetization. Although
 no magnetization is induced in the $z$-direction, it follows from Eq. (\ref{joenb}) that
this field will induce a magnetization in the radial direction since $u^r_0 = - u_{\phi}
( q/ m) \varepsilon_{\phi r z} E_z$. This is  finite since the matter phase current
$u_{\phi} \neq 0$.

 From these simple examples it is
clear that the non-Abelian nature of the mesoscopic spin transport underlying the AC
effect renders it to be a much less robust affair than its Abelian Aharonov Bohm
 counterpart. In the standard treatment these subtleties are worked under the rug and it
would be quite worthwhile to revisit this physics in detail, both experimentally and
theoretically, to find out if there are further surprises. This is however not the aim of this
paper. The general message is that even in this rather well behaved mesoscopic regime
 already finds the first signs of the fragility of non-Abelian transport. On the one hand, this
 will turn out to become lethal in the classical regime, while on the other hand we will 
demonstrate that
the coherent transport structures highlighted in this section will acquire hydrodynamical
robustness when combined with the rigidity of non-Abelian superfluid order.

\section{Spin  currents are only covariantly conserved.} \label{spincurrcovcons}

It might seem odd that the quantum equations of the previous section did not have any
resemblance to a continuity equation associated with the conservation of spin density.
To make further progress in our pursuit to describe macroscopic spin hydrodynamics
an equation of this kind is required, and it is actually straightforward to derive using a
different strategy (see also Jackiw et al\cite{jackiw1,jackiw2}).

Let us define a spin density operator,
    \beq
      \Sigma^a = \rho S^a
    \eneq
and a spin current operator,
    \begin{align}
      \begin{aligned}
    \vec j^a &= - \frac{i\hbar}{2m}\left[\psi^\dagger
    \frac{\tau^a}{2} \nabla\psi - (\nabla\psi)^\dagger
    \frac{\tau^a}{2} \psi \right] \\
    &\equiv \, \vec j_{NC}^a \quad + \vec j_{C}^a \; .
    \label{spincurr}
      \end{aligned}
    \end{align}
We observe that the spin current operator can be written as a sum of two contributions.
The first piece can be written as
    \beq
      \vec j^a_{NC} = \rho \vec u S^a \,.
    \eneq
 It factors in the phase velocity associated with the Abelian mass current $\vec u$ times
the non-Abelian charge/spin density $\Sigma^a$ carried around by the mass current.
This 'non-coherent' (relative to spin) current is according to the simple classical intuition
of what a spin current is: particles flow with a velocity $\vec{u}$ and every particle
carries around a spin. The less intuitive, 'coherent' contribution to the spin current
 needs entanglement of the spins,
    \beq
      \vec j^a_{C} = \frac{\rho}{2} \vec u^b \{S^a,S^b\} =
      \frac{\rho}{4} \vec u^a
    \eneq
and this is just the current associated with the non-Abelian phase velocity $\vec u^a$
already highlighted in the previous section.

The above expressions for the non-Abelian currents are of relevance to the 'neutral'
spin fluids, but we have to deal with the gauged currents, for instance because of
SO-coupling. Obviously we have to substitute covariant derivatives for the normal
derivatives,
     \begin{align}
      \vec J^a &= - \frac{i\hbar}{2m}\left[\psi^\dagger
      \frac{\tau^a}{2} \vec D\psi - (\vec D\psi)^\dagger
      \frac{\tau^b}{2} \psi \right] \\
      \nonumber &= \vec J S^a + \frac{\rho}{4} \left(\vec u^a -
      \frac{q}{m}\vec A^a\right) \\
      \label{covspincurr}
      &\equiv \vec J_{NC}^a \;\; + \quad \vec J_{C}^a\; ,
    \end{align}
where the gauged version of the non-coherent  and coherent currents are respectively,
  \begin{align}
  J_{NC}^a &= \vec J S^a  \label{nccovcur} \\
J_{C}^a &= \frac{\rho}{4} \left(\vec u^a -  \frac{q}{m}\vec A^a\right)
\label{ccovcur}
   \end{align}
with the Abelian (mass) current $\vec J$ given by Eq. (\ref{curra}).

It is a textbook exercise to demonstrate that the following 'continuity' equations holds
for  a Hamiltonian characterized by  covariant derivatives (like the Pauli Hamiltonian),
    \beq D_0
      \Sigma^a + \vec D \cdot\vec J^a = 0\, .
      \label{covconslaw}
    \eneq
with the usual non-Abelian covariant derivatives of vector-fields,
    \beq
      D_\mu B^a = \partial_\mu B^a + \frac{q}{\hbar}\epsilon^{abc}
      A_\mu^b B^c \, .
    \eneq
 Eq. (\ref{covconslaw}) has the structure of a continuity equation, except that the
derivatives are replaced by covariant derivatives. It is well
known\cite{weinbergvol2ch2} that in the non-Abelian case such covariant
'conservation' laws fall short of being real conservation laws of the kind encountered in
the Abelian theory. Although they impose a local continuity, they fail with regard
 to global conservation because they do not correspond with total derivatives. This is
easily seen by rewriting Eq. (\ref{covconslaw}) as
    \beq
      \partial_0\Sigma^a + \nabla\cdot\vec J^a =
      -\frac{q}{\hbar}\epsilon^{abc}A_0^b\Sigma^c -
      \frac{q}{\hbar}\epsilon^{abc} \vec A^b\cdot\vec J^c
    \eneq

The above is standard lore. However, using the result Eq. (\ref{messphall}) from the previous
section, we can obtain a bit more insight in the special nature of the phase coherent spin
current, Eq. (\ref{ccovcur}). Eq. (\ref{messphall}) can be written in covariant form as
    \beq \label{covJi}
      \vec D \cdot \vec J_{C}^a = 0 \; ,
    \eneq
involving only the space components and therefore
    \beq \label{covSigma}
      D_0\Sigma^a + \vec D \cdot \vec J_{NC}^a = 0 \, .
    \eneq
Since $\Sigma^a$ is spin density, it follows rather surprisingly that the {\em coherent
part of the spin current cannot give rise to spin accumulation}!  Spin accumulation is
entirely due to the non-coherent part of the current. Anticipating what is coming, the
currents in the spin superfluid are entirely of the coherent type and this 'non-accumulation
 theorem' stresses the rather elusive character of these spin supercurrents: they are so
'unmagnetic' in character that they are even not capable of causing magnetization when
they come to a standstill due to the presence of a barrier!

 As a caveat, from the definitions of the coherent- and non-coherent spin currents the
following equations can be derived
    \begin{align}
      \label{curljs}
      &\rho\left( \nabla\times\vec J_{NC}^a \right) =
      4\frac{m}{\hbar}\epsilon^{abc} \vec J_{C}^b\times\vec J_{NC}^c +
      \frac{q}{\hbar} \rho \epsilon^{abc} \vec A^b\times\vec J_{NC}^c
      \\
      \nonumber &\rho\left( \nabla\cdot\vec J_{NC}^a \right) \;\, =
      -\frac{1}{2}\frac{\partial\rho^2}{\partial t}S^a
      -4\frac{m}{\hbar}\epsilon^{abc}\vec J_{C}^b\cdot\vec J_{NC}^c \\
      \label{divjs}
      & \qquad\qquad\qquad\, - \frac{q}{\hbar} \rho \epsilon^{abc}
      \vec A^b\cdot\vec J_{NC}^c \, .
    \end{align}
From these equations it follows that the coherent currents actually do influence the way
that the incoherent currents do accumulate magnetization, but only indirectly. Similarly,
using the divergence of the Abelian covariant spin current together with the covariant
conservation law, we obtain the time rate of precession of the local spin density
\beq
      \partial_0\Sigma^a = \frac{\partial\rho}{\partial t}S^a +
      4\frac{m}{\hbar\rho}\epsilon^{abc}\vec J_{C}^b\cdot\vec J_{NC}^c -
      \frac{q}{\hbar}\epsilon^{abc} A_0^b \Sigma^c \, .
   \eneq
demonstrating that this is influenced by the presence of coherent- and incoherent
currents flowing in orthogonal non-Abelian directions.

This equation forms the starting point of the discussion of the (lack of) hydrodynamics of the
classical non-Abelian/spin fluid.

\section{Particle based non-Abelian hydrodynamics, or the classical spin fluid.}

We have now arrived at a point that we can start to address the core-business of
this paper: what can be said about the collective flow properties of large assemblies
of interacting particles carrying spin or either non-Abelian charge? In other words,
what is the meaning of spin- or non-Abelian hydrodynamics? The answer is: if there
is no order-parameter protecting the non-Abelian phase coherence on macroscopic
scales {\em spin flow is non-hydrodynamical}, i.e. macroscopic flow of spins does not
even exist.

The absence of order parameter rigidity means that we are considering classical spin fluids as
they are realized at higher temperatures, i.e. away from the mesoscopic regime of the previous
section and the superfluids addressed in Section \ref{spinhydro}. The lack of hydrodynamics is
well understood in the spintronics community: after generating a spin current is just
disappears after a time called the spin-relaxation time. This time depends of the effective
spin-orbit coupling strength in the material but it will not exceed in even the most favorable
cases the nanosecond regime, or the micron length scale. Surely, this is a major (if not
fundamental) obstacle for the use of spin currents for electronic switching purposes. Although
spin currents are intrinsically less dissipative than electrical currents it takes a lot of
energy to replenish these currents, rendering spintronic circuitry as rather useless as
competitors for Intel chips.

Although this problem seems not to be widely known in corporate head quarters, or either
government funding agencies, it is well understood in the scientific community. This seems to
be a different story in the community devoted to the understanding of the quark-gluon plasma's
produced at the heavy ion collider at Brookhaven.  In these collisions a 'non-Abelian fire
ball' is generated, governed by high temperature quark-gluon dynamics: the temperatures
reached in these fireballs exceed the confinement scale. To understand what is happening one
of course needs a hydrodynamical description where especially the fate of color (non-Abelian)
currents is important. It seems that the theoretical mainstream in this pursuit is preoccupied
by constructing Boltzmann type transport equations. Remarkably, it does not seem to be widely
understood that one first needs a hydrodynamical description, before one can attempt to
calculate the numbers governing the hydrodynamics from microscopic principle by employing
kinetic equations (quite questionable by itself given the strongly interacting nature of the
quark-gluon plasma). The description of the color currents in the quark-gluon plasma is
suffering from a fatal flaw: {\em because of the lack of a hydrodynamical conservation law
there is no hydrodynamical description of color transport.}

The above statements are not at all original in this regard: this case is forcefully made in
the work by Jackiw and coworkers \cite{jackiw1, jackiw2} dealing with non-Abelian 'hydrodynamics'. 
It might
be less obvious, however, that precisely the same physical principles are at work in the
spin-currents of spintronics: spintronics can be viewed in this regard as 'analogous
system' for the study of the dynamics of quark-gluon plasma's. The reason for the
analogy to be precise is that the reasons for the failure of hydrodynamics reside in the
parallel transport structure of the matter fields, and the fact that the 'gauge fields' of
spintronics are in 'fixed frame' is irrelevant for this particular issue.

The discussion by Jackiw et al. of classical ('particle based') non-Abelian 'hydrodynamics'
starts with the covariant conservation law we re-derived in the previous section, Eq.
(\ref{covSigma}). This is still a microscopic equation describing the quantum physics of a
single particle and a coarse graining procedure has to be specified in order to arrive at a
macroscopic continuity equation. Resting on the knowledge about the Abelian case this coarse
graining procedure is unambiguous when we are interested in the (effective) high temperature
limit. The novelty as compared the Abelian case is the existence of the coherent current
$\vec{J}^a_C$ expressing the transport of the {\em entanglement} associated with non-Abelian
character of the charge; Abelian theory is special in this regard because there is no room for
this kind of entanglement. By definition, in the classical limit quantum entanglement cannot
be transported over macroscopic distances and this implies that the expectation value $\langle
\vec{J}^a_C \rangle$ cannot enter the macroscopic fluid equations. Although not stated
explicitly by Jackiw et al, this particular physical assumption (or definition) is the crucial
piece for what follows -- the coherent current will acquire (quantum) hydrodynamic status when
protected by the order parameter in the spin-superfluids.

What remains is the non-coherent part, governed by the pseudo-continuity equation Eq.
(\ref{covSigma}). Let us first consider the case that the non-Abelian fields are absent (e.g.,
no spin-orbit coupling) and the hydrodynamical status of the equation is immediately obvious
through the Ehrenfest theorem. The quantity $\Sigma^a \rightarrow \langle \rho S^a \rangle$
becomes just the macroscopic magnetization (or non-Abelian charge density) that can be written
as $ n \vec{Q}$, i.e. the macroscopic particle density $n = \langle \rho \rangle$ times their
average spin $\vec{Q} = \langle \vec S \rangle$. Similarly, the Abelian phase current $\rho
\vec u$ turns into the hydrodynamical current $n \vec v$ where $\vec v$ is the velocity
associated with the macroscopic 'element of fluid'. In terms of these macroscopic quantities,
the l.h.s. of Eq. (\ref{joenb}) just expresses the hydrodynamical conservation of uniform
magnetization in the absence of spin-orbit coupling. In the presence of spin orbit coupling
(or  gluons) the r.h.s. is no longer zero and, henceforth, uniform magnetization/color charge
is no longer conserved.

Upon inserting these expectation values in Eq.'s (\ref{consn}), (\ref{covSigma}) one obtains
the equations governing classical non-Abelian fluid flow,
\begin{eqnarray}
\partial_t n + \nabla \cdot ( n \vec{v} )  & = & 0 \label{masscont} \\
\partial_t Q^a + \vec{v} \cdot \nabla Q^a & = & - \varepsilon_{abc} \left( c A^0_b
+ \vec{v} \cdot
\vec{A}^b \right) Q^c \label{nachargedyn}
\end{eqnarray}
Eq. (\ref{masscont}) expresses the usual continuity equation associated with (Abelian) mass
density. Eq. (\ref{nachargedyn}) is the novelty, reflecting the non-Abelian parallel transport
structure, rendering the substantial time derivative of the magnetiziation/color charge to
become dependent on the color charge itself in the presence of the non-Abelian gauge fields.
To obtain a full set of hydrodynamical equations, one needs in addition a 'force'
(Navier-Stokes) equation expressing how the Abelian current $n \vec{v}$ accelerates in the
presence of external forces, viscosity, etcetera. For our present purposes, this is of
secondary interest and we refer to Jackiw {\em et al.}\cite{jackiw1,jackiw2}for its form in the case of a
perfect (Euler) Yang-Mills fluid.

Jackiw{\em et al.} coined the name 'Fluid-Wong Equations' for this set of equations governing
classical non-Abelian fluid flow. These would describe a hydrodynamics that would be
qualitatively similar to the usual Abelian magneto-hydrodynamics associated with
electromagnetic plasma's were it not for Eq. (\ref{nachargedyn}): this expression shows
that the color charge becomes itself
dependent on the flow. This unpleasant fact renders the non-Abelian flow to become
non-hydrodynamical.

We perceive it as quite instructive to consider what this means in the spintronics
interpretation of the above.  Translating the gauge fields into the physical electromagnetic
fields of the Pauli equation, Eq. (\ref{nachargedyn}) becomes,
\begin{equation}
\partial_t Q^a + \vec{v} \cdot \nabla Q^a =  \left( \left[ c \vec B +  \vec{v}
\times
\vec{E} \right] \times \vec{Q} \right)_a
\label{preceseq}
\end{equation}
where $\vec{Q}(\vec{r})$ has now the interpretation of the uniform magnetization
associated with the fluid element at position $\vec{r}$. The first term on the  r.h.s.
is just expressing that the magnetization will have a precession rate in the comoving
frame, proportional to the external magnetic field $\vec{B}$. However, in the presence of
spin-orbit coupling (second term) this rate will also become dependent on the velocity of
the fluid element itself when an electrical field $\vec{E}$ is present with a component at
a right angle both to the direction of the velocity $\vec v$ and the magnetization itself.
This velocity dependence wrecks the hydrodynamics.

The standard treatments in terms of Boltzmann equations lay much emphasis on quenched
disorder, destroying  momentum conservation. To an extent this is obscuring the real  issues,
and let us instead focus on the truly hydrodynamical flows associated with the Galilean
continuum. For a given hydrodynamical flow pattern, electromagnetic field configuration and
initial configuration of the magnetization, Eq. (\ref{preceseq}) determines the evolution of
the magnetization. Let us consider two elementary examples. In both cases we consider a
Rashba-like\cite{rashba} electromagnetic field configuration: consider flow patterns in the
$xy$ directions and a uniform electrical field along the $z$ direction while $\vec{B} = 0$.

{\em a. Laminar flow}

\begin{figure}[ht!]
      \centering
      \rotatebox{0}{
    \resizebox{7.3cm}{!}{%
      \includegraphics*{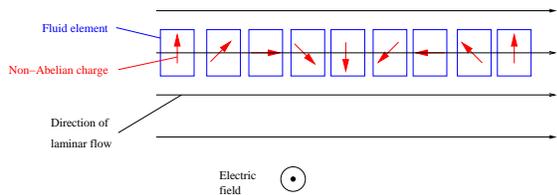}}}
      \caption{ Laminar flow of a classical spin fluid in an electric field. The fluid
               elements (blue) carry non-Abelian charge, the red arrows indicating the spin
               direction. The flow
               lines are directed to the right, and the electric field is pointing outwards
               of the paper. Due to Eq. (\ref{preceseq}), the spin precesses as indicated. }
             \label{laminarspinflow}
    \end{figure}

\begin{figure}[ht!]
      \centering
      \rotatebox{0}{
    \resizebox{7.3cm}{!}{%
      \includegraphics*{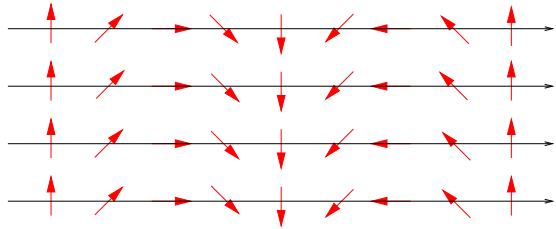}}}
      \caption{The laminar flow of a parallel transported spin current, Figure 
               \ref{laminarspinflow}, can also be viewed as a static spin spiral magnet.}
              \label{statspinspiral}
    \end{figure}

Consider a smooth, non-turbulent laminar flow pattern in a 'spin-fluid tube' realized under
the condition that the  Reynold's number associated with the mass flow is small. Imagine that
the fluid elements entering the tube on the far left  have their magnetization $\vec Q$
oriented in the same direction (Fig. \ref{laminarspinflow}). 
Assume first that the velocity $\vec v$ is uniform
inside the tube and it follows directly from Eq. (\ref{preceseq}) that the $\vec{Q}$'s will
precess with a uniform rate when the fluid elements move trough the tube. Assuming that the
fluid elements arriving at the entry of the tube have the same orientation at all times, the
result is that an observer in the lab frame will measure a static 'spin spiral' in the tube,
see Fig. \ref{statspinspiral}.  At first sight this looks like the spiral spin structures responsible 
for the ferroelectricity in the multiferroics but this is actually misleading: as we will see in Section
VII these are actually associated with localized particles (i.e. no Abelian flow) while they are rooted
instead in the entanglement current. We leave it as an excercise for the reader to demonstrate
 that the spiral pattern actually will not change when the flow in the tube acquires a  typical laminar, 
 non-uniform velocity distribution, with the velocities vanishing at the walls.

 {\em b. Turbulent flow}

\begin{figure}[ht!]
      \centering
      \rotatebox{0}{
    \resizebox{7.3cm}{!}{%
      \includegraphics*{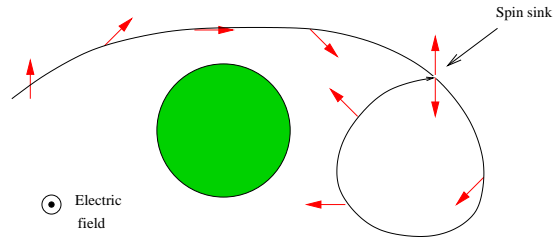}}}
      \caption{ Turbulent spin flow around an obstruction in an electric field. It is seen
                that only the ``mass'' is conserved. The change in spin direction after one
                precession around the obstruction causes a spin sink. Hence it is precisely 
                the parallel transport, or the covariant conservation, which destroys 
                hydrodynamic conservation for non-Abelian charge.   }
               \label{turbspinflow}
    \end{figure}

 Let us now consider the case that the fluid is moving much faster, such that
downstream of an obstruction in the flow turbulence arises in the matter current. In
Figure \ref{turbspinflow} we have indicated a typical stream line showing that the flow is now
characterized by a finite vorticity in the region behind the obstruction. Let us now repeat
the exercise, assuming that fluid elements arrive at the obstruction with aligned
magnetization vectors. Following a fluid element when it traverses the region with finite
circulation it is immediately obvious that even for a fixed precession rate {\em the
non-Abelian charge/magnetization becomes multivalued when it has travelled around the
vortex!} Henceforth, at long times the magnetization will average away and the spin
current actually disappears at the 'sink' associated with the rotational Abelian flow. This
elementary example highlights the essence of the problem dealing with non-Abelian
'hydrodynamics': the covariant conservation principle underlying everything is good
enough to ensure a {\em local} conservation of non-Abelian charge so that one
 can reliably predict how the spin current evolves over infinitesimal times and distances.
 However, it fails to impose a {\em global} conservation. This is neatly illustrated in this
 simple hydrodynamical example: at the moment the mass flow becomes topologically
 non-trivial it is no longer possible to construct globally consistent non-Abelian flow
 patterns with the consequence that the spin currents just disappear.

 Although obscured by irrelevant details, the above motive has been recognized in the
 literature on spin flow in semiconductors where it is known as D'yakonov-Perel spin
 relaxation\cite{dprelax}, responsible for the longitudinal ($T_1$) spin relaxation time.
 We hope that the analogy with spin-transport in solids is helpful for the
 community that is trying to find out what is actually going on in the quark-gluon fireballs.
 Because one has to deal eventually with the absence of hydrodynamics we are
pessimistic with regard to the possibility that an {\em elegant} description will be found, in
a way mirroring the state of spintronics.  We will instead continue now with our
exposition of the remarkable fact that the rigidity associated with order parameters is
not only simplifying the hydrodynamics (as in the Abelian case) but even making it
possible for hydrodynamics to exist!

\section{Electrodynamics of Spin-Orbit Coupled Systems}

Before we address the interesting and novel effects in multiferroics and spin superfluids,
we pause to obtain the electrodynamics of spin orbit coupled systems. From the Pauli Maxwell
Lagrangian (\ref{lag}) we see that the spin current couples directly to the electric field and
will thus act as a source for electric fields. In order to see how this comes about let us
obtain the electrodynamics of a spin-orbit coupled system. We presuppose the usual definition
of electromagnetic fields in terms of gauge potentials, which implies the Maxwell equations
    \beq
      \nabla\cdot\vec B = 0 \, , \qquad \nabla\times\vec E +
      \partial_0\vec B = 0 \, .
    \eneq
If we vary the Lagrangian with respect to the scalar electromagnetic potential, we obtain
    \beq
      \label{elecfield}
      \partial_iE_i = 4\pi q\epsilon_{ial}\left(
      \chi^\dagger\partial_i J_l^a \chi \right)
    \eneq
where we suppose that the charge sources are cancelled by the background ionic lattice
of the material or that we have a neutral system. This term is extremely interesting
because it says that the ``curl'' of spin
currents are sources for electric fields. In fact, the electric field equation is nothing but
the usual Maxwell equation for the electric displacement $\nabla\cdot\vec D = 0$ where
$\vec D= \vec E + 4\pi \vec P$ with
    \beq
      P_i = - \epsilon_{ial}\chi^\dagger J_l^a \chi \; .
    \eneq
The spin current acts as an electrical polarization for the material. The physical origin of this
polarization is relativistic. In the local frame the moving spins in the current produce a
magnetic field as they are magnetic moments.  After a Lorentz transformation to the lab frame,
part of this field becomes electric.  On the other hand, it can be shown that $\nabla \cdot
\vec P =0$ unless the spin current has singularities. Thus, in the absence of singularities
spin currents cannot create electric fields.

Varying the Lagrangian (\ref{lag}) with respect to the vector potential we obtain
    \begin{align}
      \begin{aligned}
    \left(\nabla\times\vec B\right)_i &= 4\pi\vec J_{em} - 4\pi
    \left(\nabla\times q\vec\Sigma\right)_i + \partial_0E_i \\
    &- 4\pi q\epsilon_{lai}\partial_0\left(\chi^\dagger j_l^a \chi
    \right) \\
    &= 4\pi\vec J_{em} - 4\pi \left(\nabla\times
    q\vec\Sigma\right)_i + \partial_0D_i \; .
      \end{aligned}
    \end{align}
The first term on the right hand side contains the usual electromagnetic current
    \beq
      \vec J_{em} = 4\pi e\rho\left(u_i + u_i^a\chi^\dagger
      S^a\chi\right)
    \eneq
which includes the motion of particles due to the advance of the Abelian and the
non-Abelian
phases. The term containing the non-Abelian velocity (the coherent spin current) in this
electromagnetic current will only contribute when there is magnetic order
$\langle S^a \rangle \neq 0$. The second term is conventional since it is the curl of the
magnetization which generates magnetic fields. The
third is the Maxwell displacement current in accordance with our identification of the
electrical polarization caused by the spin current.

 \section{Spin hydrodynamics rising from the ashes I: the spiral magnets.}

 Recently the research in multiferroics has revived. This refers to materials that are at the
 same time ferroelectric and ferromagnetic, while both order parameters are coupled.
The physics underlying this phenomenon goes back to the days of Lifshitz and
Landau\cite{continuumel}. Just from considerations regarding the allowed invariants in
the free energy it is straightforward to find out that when a crystals lacks an inversion
center (i.e., there is a net internal electric field) spin-spin interactions should exist giving
rise to a spiral modulation of the spins (helicoidal magnets). The modern twist of this
argument is\cite{mostovoy}: the spin spiral can be caused by magnetic frustration as well,
 and it now acts as a cause (instead of effect) for an induced ferroelectric polarization.
Regarding the microscopic origin of these effects, two mechanisms have been identified. The
first one is called 'exchange striction' and is based on the idea that spin-phonon
interactions of the kind familiar from spin-Peierls physics give rise to a deformation of the
crystal structure when the spin-spiral order is present, and these can break inversion
symmetry\cite{cheong}. The second mechanism  is of direct relevance to the present
subject matter. As we already explained in the previous section, a spiral in the spin-density
can be viewed at the same time as a spin current. In the presence of the
magnetic order parameter this spin current acquires rigidity (like a supercurrent) and
therefore it can impose its will on the 'gauge' fields. In the spin-orbital coupling case, the
'gauge' field of relevance is the physical electrical field, and henceforth the 'automatic'
spin currents associated with the spiral magnet induce an electrical field via the spin-orbit
coupling, rendering the substance to become a ferroelectric\cite{nagaosabalatskii}.

This substance matter is rather well understood \cite{mostovoy} and the primary aim of this
section is to explain how these 'spiral magnet' spin currents fit into the greater picture of
spin-hydrodynamics in general. Viewed from this general perspective they are quite
interesting: they belong to a category of non-Abelian hydrodynamical phenomena having no analogy
in the Abelian universe. On the one hand these currents are spontaneous and truly
non-dissipative and in this regard they are like Abelian supercurrents. They should not be
confused with the Froehlich 'super' currents associated with (Abelian) charge density waves:
these require a time dependence of the density order parameter (i.e., the density wave is
sliding) while the spiral magnet currents flow also when the non-Abelian density (the spiral)
is static. This belies their origin in the coherent non-Abelian phase current $\vec{J}^a_{C}$ 
just as in the spin-superfluids, or either the non-Abelian Higgs phase. 

An important property of the static coherent spin 
currents of the spin spirals is that vortex textures in the spin background  become sources
of electrical charge in the presence of spin-orbit coupling, as first observed by Mostovoy \cite{mostovoy} .  
Anticipating the discussion of the SO coupled spin superfluid in the next sections, a major difference between
those and the multiferroics is that in the former the phase coherent spin fluid can {\em quantize} the electrical line charge but not {\em cause} electrical charge because of the
important difference that such a current can not originate spontaneously in the spin superfluid 
because it needs to be created by an electric field. It can trap charge because being a supercurrent
it does not decay if the battery that creates the electric field is removed.

Last but not least, the spiral magnet currents offer a minimal context to illustrate the most
fundamental feature of non-Abelian hydrodynamics: the rigidity of the order parameter is
capable of restoring hydrodynamical degrees of freedom that are absent in the 'normal'
fluid at high temperature. This is so simple that we can explain it in one sentence. One
directly recognizes the XY spin vortex in the turbulent flow of Fig. \ref{turbspinflow}, 
but in the presence
of spin density order the 'spiral' spin pattern associated with the vortex has to be single
valued, and this in turns renders the spin current to be single valued: spin currents do not
get lost in the ordered magnet!

To become more explicit, let us rederive Mostovoy's result in the language of this paper, by
considering  an ordered $XY$-magnet with an order parameter that is
the expectation value of the local spin operator
    \beq
      \langle S_x + i S_y\rangle = S e^{i\theta}
    \eneq
 In general a spin state of an $XY$-magnet is given by
    \beq
      \prod_\text{lattice sites} g (\vec x) |\uparrow \rangle
    \eneq
where we specialize to spin 1/2  for explicitness, but similar results hold for larger
spin. $|\uparrow\rangle$ is a spinor in the $+z$ direction and $g(\vec x)$ is an $SU(2)$
rotation matrix in the $xy$-plane:
    \beq
      g(\vec x) = e^{i\theta (\vec x) \tau_z/2}
    \eneq
where $\tau_z$ is the Pauli matrix in the $z$-direction. The  ordered ground
state of the uniform  $XY$-magnet requires that $\theta (\vec x)$ and hence $g(\vec x)$ are 
independent of $\vec x$. Besides the ground state, $XY$-magnets have
excited metastable states corresponding to XY spin vortices. These are
easily constructed by choosing
    \beq
      \theta(\vec x) = n \phi \;, \quad n \, \text{integer} \;, \quad
      \phi = \arctan \left(\frac{y}{x}\right) \,.
    \eneq
Now we can compute the spin current in this state. The coherent spin current is
given by
    \beq
      \vec J^a_C = \frac{\hbar\rho}{2m} \vec u^a = -i
      \frac{\hbar\rho}{2m} \left[ g^{-1} \frac{\tau^a}{2} \nabla g -
      (\nabla g^{-1}) \frac{\tau^a}{2} g \right] \,.
    \eneq
For our case
    \begin{align}
      \begin{aligned}
    g^{-1} \frac{\tau_x}{2} g &= \frac{1}{2} \left[ \tau_x
    \cos\theta + \tau_y \sin\theta \right] \\
    g^{-1} \frac{\tau_y}{2} g &= \frac{1}{2} \left[ -\tau_x
    \sin\theta + \tau_y \cos\theta \right]
      \end{aligned}
    \end{align}
we have the appropriate $O(2)$ or $U(1)$ rotation. We also have for
the vortex $\theta = n\varphi$
    \begin{align}
      \begin{aligned}
    J^a_c &= \frac{n\hbar\rho}{8m} \nabla\varphi \left[ e^{-i n
    \varphi \tau^z/2} \left\{ \tau^a,\tau^z\right\} e^{i n \varphi
    \tau^z/2} \right] \\
    &= \frac{n\hbar\rho}{4m} (\nabla \varphi) \delta^{az} \,.
      \end{aligned}
    \end{align}
According to the results in the previous section, spin currents alter the electrodynamics via  
 Gauss' law,
    \beq
      \partial_i E_i = 4\pi q \epsilon_{ial} \langle \partial_i J_l^a
      \rangle
    \eneq
where $q$ measures the coupling between spin currents and electric
fields via spin orbit coupling. Hence, using that for $\phi = \arctan
(y/x)$,
    \beq
      \nabla \times \nabla\phi = 2\pi \delta^{(2)} (\vec r)
    \eneq
we find for the spin current of the vortex,
    \beq
      \partial_i E_i = 2 \pi^2 n q \frac{\hbar\rho}{m} \delta^{(2)}
      (\vec r) \,.
    \eneq
Therefore spin vortices in $XY$-magnets produce electric fields!

\section{Spin hydrodynamics rising from the ashes II: the spin superfluids} \label{spinhydro}

Even without knowing a proper physical example of a spin-orbit coupled spin-superfluid
one can construct its order parameter theory using the general principles discovered by
Ginzburg and Landau. One imagines a condensate formed from electrically neutron
bosons carrying $SU(2)$ spin triplet quantum numbers. This condensate is
characterized by a spinorial order parameter,
  \beq
      \Psi = | \Psi | \; e^{(i\theta + i\varphi^a\tau^a/2)} \chi
    \eneq
 where $ | \Psi |$ is the order parameter amplitude, nonzero in the superfluid state, while
$\theta$ is the usual $U(1)$ phase associated with number, while the three non-Abelian phases
$\varphi^a$, with the Pauli matrices $\tau^a$ acting on a reference spinor $\chi$ keep track
of the $SU(2)$ phase structure. According to the Ginzburg-Landau recipe, the free energy of
the system should be composed of scalars constructed from $\Psi$, while the {\em gradient
structure should be of the same covariant form as for the microscopic problem} -- parallel
transport is marginal under renormalization. Henceforth, we can directly write down the
Ginzburg-Landau free energy density for the spin superfluid in the presence of spin orbit
coupling,
    \begin{align}
      \begin{aligned}
     \mathcal F  &= i \hbar  \psi^\dagger D_0 \psi +
    \psi^\dagger\frac{\hbar^2}{2m}\vec D^2\psi  + m^2 |\Psi|^2 \\
    & + w |\Psi|^4 + \frac{1}{2m}\psi^\dagger \frac{q^2}{4}\vec A^a
    \cdot\vec A^a \psi \\
    \label{covlag}&+ \frac{1}{8\pi} \; \left( E^2 -B^2 \right) \, .
      \end{aligned}
    \end{align}
 We now
specialize
to the deeply non-relativistic case where the time derivatives can be ignored, while we
consider electrically neutral particles ($e = 0$) so that the EM gauge fields drop out
from the covariant derivatives.

Well below the superfluid transition the amplitude $| \Psi |$ is finite and frozen and one can
construct a London-type action. Using the formulas in the appendix we obtain that
 \beq
      \label{lsv}
      \mathcal L_{\text{spin-vel}} = -\frac{m}{8}\rho \left( \vec u^a
      - \frac{m}{2} \rho \vec u^2 - \frac{q}{m}\vec A^a \right)^2 + \frac{q^2}{8m}\vec
      A^a\cdot\vec A^a \, .
    \eneq
Using the spin identities defined in Section \ref{spincurrcovcons}, this can be rewritten as
 \beq
      \label{lsv2}
      \mathcal L_{\text{spin-vel}} = -2 \vec J_C^a \cdot \vec J_C^a - 2
    \vec J_{NC}^a \cdot \vec J_{NC}^a
    - \frac{q}{m}\left( \vec A^a \right)^2 + \frac{q^2}{8m}\vec
      A^a\cdot\vec A^a \, .
    \eneq
We see that the Ginzburg-Landau action is a sum of the spin coherent and non-coherent squared
currents. The spin noncoherent part has to do with mass or $U(1)$ currents, but since the
particles carry spin they provide a spin current only if $\langle S^a \rangle \neq 0 $, requiring a net
magnetization. The coherent part is a bona fide spin current originating
in the coherent advance of the non-Abelian phase associated with the spin direction.

In order to make contact with the Helium literature\cite{minvol} we will write our spin operators and the
coherent spin currents in terms of $SO(3)$ rotation matrices via
    \beq
      R^a_{\;b}(\vec \varphi) \; \frac{\tau^b}{2} =
      e^{-i\varphi^a\tau^a/2} \; \frac{\tau^a}{2} \;
      e^{i\varphi^a\tau^a/2}
    \eneq
with $R^a_{\;b}(\vec \varphi)$ an $SO(3)$ rotation matrix around the vector
$\vec \varphi$ by an angle $|\vec \varphi|$, we obtain that the spin operator is a local
$SO(3)$ rotation of the Pauli matrices
    \beq
      S^a=R^a_{\;b}(\vec \varphi)\; \frac{\tau^b}{2}\; \, .
    \eneq
In terms of the rotation operators, the spin velocities related to advance of the
non-Abelian phase are
    \beq
      \label{uamat}
      \vec u^a= \frac{\hbar}{m} \epsilon_{abc} [\nabla R^b_{\;d}(\vec
      \varphi)] R^d_{\;c}(\vec \varphi) \, .
    \eneq
It is also easily seen that
    \beq
      \label{u0mat}
      u_0^a = \epsilon_{abc} [\partial_0 R^b_{\;d}(\vec \varphi)]
      R^d_{\;c}(\vec \varphi) \, .
    \eneq

If we look at the expressions for $\vec u^a$ and $u_0^a$ in terms of the spin rotation
matrix for the spin-orbit coupled spin superfluid, Eq.'s (\ref{uamat}, \ref{u0mat}), we recognize these
to be the exact analogues of the spin velocity and spin angular velocity of $^3$He-B
(\ref{omegab}) reproduced in Section \ref{heb}. We define $g$ through
    \begin{align}
      \begin{aligned}
    R_{\alpha i}(\vec \varphi) \; \frac{\tau^i}{2} &=
    e^{-i\varphi^a\tau^a/2} \; \frac{\tau_\alpha}{2} \;
    e^{i\varphi^a\tau^a/2} \\
    &= g^{-1} \frac{\tau_\alpha}{2} g = S_\alpha \;,
      \end{aligned}
    \end{align}
that is
    \beq
      g = e^{i\varphi^a\tau^a/2} \;,
    \eneq
which is an SU(2) group element. We now have the spin velocities and angular velocities
expressed as
    \begin{align}
      \begin{aligned} \label{omegai}
    \omega_{\alpha i} &= -i \text{Tr} \left\{S_\alpha
    g^{-1}\partial_i g \right\} = -i \text{Tr} \left\{
    g^{-1}\frac{\tau_\alpha}{2}\partial_i g \right\} \\
    \omega_{\alpha} &= -i \text{Tr} \left\{S_\alpha
    g^{-1}\partial_0 g \right\} = -i \text{Tr} \left\{
    g^{-1}\frac{\tau_\alpha}{2}\partial_0 g \right\}
      \end{aligned}
    \end{align}
The first is proportional to the coherent spin current and the second to the effective
magnetization. If we define the spin superfluid density via
    \beq
      \rho = \frac{1}{\gamma^2}\chi_B c^2 \;,
    \eneq
we have the following Lagrangian that describes the low energy spin physics, written in a way
that is quite analogous to that of \he-B \cite{minvol},
    \begin{widetext}
      \beq
        \label{socsflagr}
        L(\vec\varphi,\vec E,\vec B) = \frac{1}{2\gamma^2} \chi_B
        \left( \vec\omega^2 + 2 \gamma\vec\omega\cdot\vec B \right) -
        \frac{1}{2\gamma^2} \chi_B c^2 \left(\omega_{\alpha i}^2 -
        \frac{4\mu}{\hbar c}\omega_{\alpha i}\epsilon_{\alpha ik}
        E_k\right) + \frac{1}{8\pi}\left(E^2 - B^2\right) \; .
      \eneq
    \end{widetext}
From this Lagrangian  we obtain the spin equations of
motion for the spin superfluid by varying with respect to the non-Abelian phase
    \begin{align}
      \partial_0 \left[ \frac{\partial L}{\partial(\partial_0 g)}
      \right] + \partial_i \left[ \frac{\partial
      L}{\partial(\partial_i g)} \right] - \frac{\partial L}{\partial
      g} = 0 \; .
    \end{align}
We evaluate
    \begin{align}
      & \begin{aligned}
    \quad\;\; \frac{\partial L}{\partial g} &= \frac{\partial
    g^{-1}}{\partial g} \frac{\partial \omega_\alpha}{\partial
    g^{-1}} \frac{\partial L}{\partial \omega_\alpha} +
    \frac{\partial g^{-1}}{\partial g} \frac{\partial
    \omega_{\alpha i}}{\partial g^{-1}} \frac{\partial L}{\partial
    \omega_{\alpha i}} \\
    &= -i g^{-2} \frac{\tau_\alpha}{2}(\partial_0 g)
    \frac{1}{\gamma^2} \chi_B \left(\omega_\alpha + 2\gamma
    B_\alpha\right) \\
    &+ i g^{-2} \frac{\tau_\alpha}{2}(\partial_i g)
    \frac{1}{\gamma^2} \chi_B c^2 \left(\omega_{\alpha i} -
    \frac{2\mu}{\hbar c}\epsilon_{\alpha ik} E_k\right)
      \end{aligned} \\
      & \begin{aligned}
    \frac{\partial L}{\partial (\partial_0 g)} &= \frac{\partial
    \omega_\alpha}{\partial (\partial_0 g)} \frac{\partial
    L}{\partial \omega_\alpha} \\
    &= i g^{-1}\frac{\tau_\alpha}{2} \frac{1}{\gamma^2}\chi_B
    \left(\omega_\alpha + \gamma B_\alpha\right)
      \end{aligned} \\
      & \begin{aligned}
    \frac{\partial L}{\partial (\partial_i g)} &= \frac{\partial
    \omega_{\alpha i}}{\partial (\partial_i g)} \frac{\partial
    L}{\partial \omega_{\alpha i}} \\
    &= - i g^{-1}\frac{\tau_\alpha}{2} \frac{1}{\gamma^2} \chi_B
    c^2 \left(\omega_{\alpha i} - \frac{2\mu}{\hbar
    c}\epsilon_{\alpha ik} E_k\right)
      \end{aligned}
    \end{align}
which yields the rather formidable equation of motion
    \begin{align}
      \begin{aligned}
    0 &= \partial_0 \left[ i g^{-1}\frac{\tau_\alpha}{2}
    \left(\omega_\alpha + \gamma B_\alpha\right) \right] \\
    &+ \partial_i \left[- i g^{-1}\frac{\tau_\alpha}{2} c^2
    \left(\omega_{\alpha i} - \frac{2\mu}{\hbar c}\epsilon_{\alpha
    ik} E_k\right) \right] \\
    &+ i g^{-2} \frac{\tau_\alpha}{2}(\partial_0 g)
    \left(\omega_\alpha + \gamma B_\alpha\right) \\
    &- i g^{-2} \frac{\tau_\alpha}{2}(\partial_i g) c^2
    \left(\omega_{\alpha i} - \frac{2\mu}{\hbar c}\epsilon_{\alpha
    ik} E_k\right)
      \end{aligned}
    \end{align}
After some straightforward algebra this equation reduces to the fairly
simple equation
    \beq \label{eomgeneral}
      \partial_0\left(\omega_\alpha + \gamma B_\alpha\right) - c^2
      \partial_i\left(\omega_{\alpha i} - \frac{2\mu}{\hbar
      c}\epsilon_{\alpha i k} E_k\right) = 0 \; .
    \eneq
The solution of this equation of motion gives the SU(2) group element
$g$ as a function of space and time, and the spin
velocities and angular velocities can be determined.

Similarly, by varying the Lagrangian (\ref{eomgeneral}) with respect to
the electromagnetic potentials, we obtain the Maxwell equations for
the electromagnetic fields ``created'' by the spin velocities and angular
velocities.
    \beq
      \label{efieldhe}
      \partial_kE_k = 4\pi \partial_k\left(
      \frac{2c\mu}{\hbar\gamma^2}\chi_B\epsilon_{\alpha
      ik}\omega_{\alpha i} \right)
    \eneq
    \begin{align}
      \begin{aligned}
    \left(\nabla\times\vec B\right)_\alpha &= - 4\pi
    \left(\nabla\times \frac{1}{\gamma} \chi_B \omega_\alpha
    \right) \\
    &+ \partial_0 \left( E_\alpha - 4\pi
    \frac{2c\mu}{\hbar\gamma^2}\chi_B\epsilon_{\beta
    i\alpha}\omega_{\beta i} \right)
      \end{aligned}
    \end{align}

We like to draw the reader's attention to the fact that Mineev and Volovik derived these
results already in the seventies \cite{minvol} in the context of \he-B. We show here that
these hold in the general case of an $SU(2)$ spin superfluid, and will demonstrate in section
\ref{hea} that similar equations can be derived for the case of superfluid \he-A as well.

\section{Charge trapping by spin superfluids}

We now go back to the trick of charge trapping in superfluids we used previously to
wet your appetite. How does this magic trick work? At the heart of our idea lies the spin
vortex solution. Let us first briefly sketch the argument, and then prove it. The straight
wire causes an electric field of
\begin{equation}
      \vec E = \frac{\lambda}{2\pi\varepsilon_0 r}\hat{r} ,
\end{equation}
where $\hat r$ is a radial unit vector in the $xy$ plane perpendicular to the cylinder axis
$z$. The azimuthal angle is $\varphi$. We now need to determine the electric field in the
superfluid region. Because of the symmetry of the problem, this electric field will be radial.
Lets call it $E_i$. This electric field will drive a spin current, which will be a source of
electric field itself if it has a singularity that will lie on the wire because of the radial symmetry.
 The symmetry of the problem suggests that the spins will be polarized along the axis of
the cylinder. By solving the equations of motion  in the presence of an electric field and no
magnetic field, we obtain that when the  spin current and spin angular velocity satisfy the
Spin Hall relation for spin direction $\alpha=z$
    \begin{equation}
\omega_\alpha = 0, \; \; \omega_{z\varphi} =   \frac{2\mu}{\hbar c^2} E_r,
    \end{equation}
    with the magnetic moment of the He-atoms
\begin{equation}
\mu = g \frac{m_e}{m_{He}}\mu_B,
\end{equation}
whereas the other spin superfluid velocities vanish.
Since the electric fields do not depend on the $z$-coordinate and only have a radial component,
the equations of motion Eq.(\ref{eomgeneral}) are satisfied. In our case, written in cylindrical
coordinates,
    \begin{equation}
      \vec\omega_z = \frac{2\mu}{\hbar c^2} \epsilon_{zik}E_k \sim
      \hat\varphi \; .
    \end{equation}

We see that the electric field leads to a \emph{spin vortex}, i.e., $z$-polarised spins
flowing around the wire. This is nothing different from vortices in Bose superfluids induced
by rotation.  This might cause some concern as we have an $SU(2)$ superfluid while vortices are topological
defects associated with $U(1)$. Why is this spin vortex topologically stable? This has everything to do with
the fact that we are not dealing with a real gauge theory but that our 'gauge' fields are in fact physical. In a
literal sense, the topology is 'hard wired' by the fact that we have put the wire inside the cylinder: the electrical
field is forced by the experimentalist to carry a vortex topology, and this topology is via the covariant derivatives
imposed on the spin current -- were it a real (unphysical) gauge field, it has to sort this out by itself and the 
outcome would be the usual 't Hooft-Polyakov monopole. There is a neat mathematical way of saying the same
thing. Gauge theories coupled to matter are known to
mathematicians as bundle theories. One way to classify them is by using Chern classes
\cite{gockschuck, arafune}. The Chern classes do not depend on the gauge chosen, or the
configuration of the matter fields, but are a property of the bundle. The ramification is that
if the topology of the gauge field is cylindrical, the matter field has cylindrical topology
as well.

The stability of the vortex can also be checked  by demonstrating that a vortex centered on the wire,
with a spin direction parallel to this wire, does satisfy the equations of motion we derived in Section IX, while
such a solution is an energy minimum. From the Lagrangian in the previous section it follows that the momentum
conjugate to the non-Abelian phase is
    \beq
    \mathcal H = \frac{ \chi_B c^2}{2\gamma^2} \left(\omega_{\alpha i}^2 -
        \frac{4\mu}{\hbar c}\omega_{\alpha i}\epsilon_{\alpha ik}
        E_k\right) +  \frac{1}{8\pi} E^2 \, .
    \eneq
When the vortex solution ,and thereby the Spin Hall relation is valid, the energy density becomes,
    \beq \label{spinvortexlowersenergy}
    \mathcal H_{SH}= \left( \frac{1}{8\pi} - \frac{ \chi_B c^2}{\gamma^2}
    \frac{\mu^2}{\hbar^2 c^4}\right)E^2 \; .
    \eneq
If there is no vortex we have energy density
    \beq
    \mathcal H_{\text{no-vortex}} = \frac{1}{8\pi} E^2
    \eneq
which is bigger than the energy density $\mathcal H_{SH}$ corresponding to a vortex
present and thus the solution with the vortex is
favored. If we have a vortex solution and perturb around by $\delta \omega_{\alpha i}$
the energy changes by
    \beq
    \delta \mathcal H = \frac{ \chi_B c^2}{2\gamma^2}
    \left(\delta \omega_{\alpha i}\right)^2
    \eneq
which is a positive quantity and we see that the vortex solution is stable against perturbations as
 they increase the energy of the system. We can rephrase the
above reasoning in a more sophisticated way: the cylindrical topology of the fixed-frame
gauge
fields imposes the same vortex-type topology on the matter field, because of the parallel
transport structure originating from spin-orbit coupling!

The vortex topology can be
classified by winding numbers. Indeed, from the definition of the spin supercurrent in chapter
\ref{spinhydro} we have
    \begin{equation}
      \vec\omega_z = -\nabla\theta  .
    \end{equation}
Therefore the spin current must satisfy the quantization condition
    \begin{equation}\label{quantisedomega}
      \oint\vec\omega_z\cdot d\vec l = 2\pi N
    \end{equation}
when we integrate around the cylinder where $N$ is an integer. This quantisation is not
quite shocking, since any order parameter theory has this condition. However, bearing
in mind the magnetic flux trapping in superconductors, it is interesting to integrate the
spin current after substituting the spin-Hall equation. By Gauss' law, one obtains that the
very same phase velocity integral becomes
    \begin{equation}
   \oint\vec\omega_z\cdot d\vec l   2\pi \frac{e}{m_{He}} \mu_0 \lambda .
    \end{equation}
In other words, the charge density is quantised in units of
\begin{equation}
\lambda = N\lambda_{0} = N \frac{m_{He}}{\mu_0 e }  = 2.6 \times 10^{-5} C/m !.
\end{equation}
in the specific case of \he\. This is of course a very large line-charge density, and this is of course 
rooted in the fact that this quantum is 'dual'  to the tiny spin orbit coupling of helium, in the same way that
the flux quantum in superconductors is inversely proportional to the electrical charge. In he next section
we will show that this huge required electrical charge is detremental to any attempt to realize such an
experiment employing a substance like helium. 

This experiment is the rigid realisation of the Aharonov-Casher phase \cite{aharonovcasher}, for which
our application is inspired by Balatskii and Altshuler \cite{balatskiialtshuler}. The rigidity is provided
by the superfluid density, forcing the winding number to be integer. Our idea is actually the spin
superfluid analogue of the flux trapping with superconducting rings. The
quantization of magnetic flux is provided by the screening of electromagnetic fields, causing
vanishing total superconducting current. The latter, being defined covariantly,
 consists of a $U(1)$ superfluid velocity
and a gauge field. Calculating the line integral
\begin{equation}
0 = \oint J^{sc}_i dx_i = \oint \partial_i \phi - \oint A_i dx_i = 2\pi n - \Phi_{sc},
\end{equation}
leading to the flux quantisation condition. In the above argument, the gauge fields
$A_i$ have dynamics, leading to screening of the $A_i$ in the superconducting ring.

In our case, the gauge fields are fixed by the electromagnetic fields, such that there
cannot be screening effects. Still, the spin-Hall equations, which solve
the equations of motion (\ref{eomgeneral}),
lead to a vanishing superconducting current.
The gauge fields, being unscreened, play now a quite different role: these are necessary to
force the topology of the superfluid order parameter to be $U(1)$. The result
is the same: quantisation of electric flux, determined by the charge on the wire.

Charge trapping in spin superfluids and in magnets both originate from the coherent
part of the spin current. In this sense, there is not too much difference between the two
effects. On the other hand, there is a subtle, but important distinction. For  magnets
there is no need for electric fields to impose the supercurrent, since they are wired
in by the magnetic  order. In contrast in the spin superfluids,
 an electric field is necessary to create a coherent spin current since there is no
magnetisation, and in this sense the spin superfluids cannot create electrical charge,
while magnets can.

The question which surely is nagging the reader's mind, is whether one can actually
\emph{perform} our experiment. The answer is threefold. To begin with, nobody knows
of the
existence of a material exhibiting an $SU(2)$-order parameter structure. Fortunately, the
existence of two spin superfluids is well-established: $^3$He-A and $^3$He-B.
We will show that  \he-B has an order parameter structure similar to that of the pure spin
superfluid. The effect of dipolar locking will destroy the spin vortex caused by the electric
field, however, see Section (\ref{sect:diplock}).  Then we will show that \he-A has, for
subtle reasons, the wrong topology to perform our experiment. We will also demonstrate that
the small spin-orbit coupling constant forces us to use an amount of \he\ with which one can
cover Alaska, turning our experiment into a joke. In the outlook of this work, we will discuss
how the organic superconductors \cite{kanoda03,kanoda05} might meet the desired conditions.

Let us first consider the secrets of $^3$He more generally.

\section{ \he \ and order parameter structure}

As is well-known, $^3$He is a fermionic atom carrying spin $\frac{1}{2}$. In field theory,
we describe it with an operator $c_{p\alpha}$, where $p$ is momentum and $\alpha$ is
spin. In
the normal phase, it is a Fermi liquid, but for low temperatures and/or high pressures, the
He displays a BCS-like instability towards pairing. Indeed, the condensate wave function
$\Psi$ displays an order parameter which transforms under both spin and orbital angular
momentum:
\begin{equation}
\left< \Psi \right| \sum_{\mathbf{p}} \ \mathbf{p}c_{\mathbf{p}\alpha}c_{-\mathbf{p}
\beta}
\left| \Psi \right> = A_{\mu i} (i\sigma^{\mu}\sigma^{2})_{\alpha\beta},
\end{equation}
so the order parameter describes a p-wave state. The $A_{\mu i}$ carry a spatial index
$i$ and
an internal spin index $\mu$. The numbers $A_{\mu i}$
transform as a
vector under the spin rotation group $SO(3)^{S}$ acting on the index $\mu$ and the orbital
rotation group $SO(3)^{L}$ acting on the index $i$. We can reconstruct the wave function
$|\Psi>$ from the $A_{\mu i}$ as follows. First we rewrite them as a vector decomposition
 with
amplitudes $a_{kl}$ in the following way:
\begin{equation} \label{decompvev}
A_{\mu i} = \sum_{k,l} a_{kl}\lambda_{\alpha}^{k}\lambda_{i}^{l}.
\end{equation}
The $\lambda^{k,l}$ are vectors. Then the wave function in momentum space
$\Psi(\mathbf{p})=
<\mathbf{p}|\Psi>$ is the decomposition
\begin{equation}
\Psi(\mathbf{p}) = \sum_{k,l} a_{kl} Y_{\mbox{\tiny{L=1}}, k}(\mathbf{p})\
\chi_{\mbox{\tiny{S=1}} ,l}\ \ ,
\end{equation}

where $ Y_{\mbox{\tiny{L=1}}, k}$ is a triplet spherical harmonic and
$\chi_{\mbox{\tiny{S=1}}
,l}$ is a triplet spinor. This means that the order parameter has $3\times 3 \times 2$ real
degrees of freedom. Indeed, following Volovik \cite{volovikexo} and Leggett \cite{leggetthe}, there
exist two mean-field states.

The first one is an isotropic state with vanishing total angular momentum $J=L+S=0$. In
order
to have zero projection of the total spin $m_J = m_l + m_s = 0$, we have for the coefficients
in the decomposition (\ref{decompvev})
\begin{equation}
a_{+-}=a_{-+}=a_{00}=\Delta_B.
\end{equation}
This state is called the B-phase of \he , or the BW-state, after Balian and Werthamer
\cite{BW}. This means that the order parameter looks like
\begin{equation}\label{AB}
A_{\alpha i} = \Delta_B \delta_{\alpha i}.
\end{equation}
There is still a degeneracy, however. Indeed, both the spin and orbit index transform under
$SO(3)$, which leads to an order parameter manifold
\begin{equation}
R_{\alpha i} = R^L_{ij} R^S_{\alpha\beta}\delta_{\alpha i}, \mbox{\ \ \ \ or \ \ \ \ } R =
R^S
(R^L)^{-1}.
\end{equation}
So the matrix $R \in SO(3)$ labels all degenerate vacua, and describes a \textit{relative}
rotation of spin and orbital degrees of freedom. Including also the $U(1)$ phase of the matter
field, the order parameter manifold of \he-B is
\begin{equation}
G_B = SO(3)_{rel} \times U(1)_{matter}.
\end{equation}
This will be the starting point of our considerations for \he-B, in which we will often drop
the $U(1)$ matter field.

The second one is the A-phase, which has just one non-vanishing amplitude in
(\ref{decompvev}), 
\begin{equation}
a_{0+}=\sqrt{2} \Delta_A,
\end{equation}
which corresponds to a state with $m_s=0$ and $m_l=1$. The quantisation axes are
chosen along
the $\hat{z}$-axis, but this is just arbitrary. This is known as the $^3$He-A phase, or the
Anderson-Brinkman-Morel (ABM) state \cite{ABM}. The order parameter is
\begin{equation}\label{AA}
A_{\alpha i} = \Delta_A \hat{z}_\alpha ( \hat{x}_{i} + i \hat{y}_{i}).
\end{equation}
Rotations of the quantisation axis of $^3$He-A lead to the same vacuum, which tells us how to
describe the degeneracy manifold. The vector describing spin, called the $\hat{d}$-vector in
the literature \cite{leggetthe}, can be any rotation of the $\hat{z}$-axis:
\begin{equation}
\hat{d}_\alpha = R^S_{\alpha\beta}\hat{z}_{\beta}.
\end{equation}
Since only the direction counts in which the $\hat{d}$-vector points, its order parameter
manifold is the 2-sphere $S^2$. The orbital part of the order parameter is called the
$\hat{l}$ vector, which is in the "gauge" Eq. (\ref{AA}) simply $\hat{z}$. Again, the
orientation
is arbitrary, so that any rotation $R^L$ and gauge transformation $e^{i\phi}$  leads to a
correct vacuum state,
\begin{equation}
 \hat{e}^{(1)}_{i} + i \hat{e}^{(2)}_{i} = e^{i\phi}R^L_{ij}(\hat{x}_{j} +
i \hat{y}_{j}),
\end{equation}
where $\hat{l}=e^{(1)} \times e^{(2)}$ is invariant under $e^{i\phi}$. This phase
communicates with the phase of the matter field, so that the order parameter has a relative
$U(1)_{rel} = U(1)_{matter-orbital}$. For the determination of
the order parameter manifold for He-A, we need to observe that the order parameter does not
change if we perform the combined transformation $\hat{d}\rightarrow -\hat{d}$ and
$(\hat{e}^{(1)}_{i} + i \hat{e}^{(2)}_{i}) \rightarrow - (\hat{e}^{(1)}_{i} + i
\hat{e}^{(2)}_{i})$. This means that we have to divide out an extra $\mathbb{Z}_2$ degree of
freedom. In summary, the order parameter manifold for He-A is
\begin{equation}
G_A = (S_{s}^{2} \times SO(3)_{l})/\mathbb{Z}_2,
\end{equation}
where $s$ refers to the spin and $l$ to the orbit. The intricateness of the order parameter
already indicates that there is a lot of room for various kinds of topological excitations and
other interesting physics. For extensive discussions, we recommend the books of Grigory
Volovik \cite{volovikexo, volovikdrop}. What counts for us, however, is how the topology is
influenced by switching on fixed frame gauge fields.

\subsection{\he -B}\label{heb}

As discussed above, the order parameter of \he \ is described by an $SO(3)$ matrix $R$.  The
question is now if $R$ admits spin vortex solutions. In principle, it does, because $SU(2)$
rotations are like $SO(3)$ rotations, since they are both representations of angular momentum,
as we learned in freshman quantum mechanics courses. This means that, in principle, all
considerations for the $SU(2)$ case apply to \he-B as well. In particular, the spin superfluid
velocity Eq. (\ref{omegai}) has a similar expression, but now with $ g = R \in SO(3)$. It reads
\begin{equation}\label{omegab}
\omega_{\alpha i} = \frac{1}{2}
      \epsilon_{\alpha\beta\gamma}R_{\beta j} \partial_i R_{\gamma j}.
\end{equation}

 Inspired by the $SU(2)$
case, which was effectively Abelianised, we try a vortex solution around the $z$-axis
(assuming the electric field is radial)
\begin{equation} \label{u1R}
 R = \exp(i\theta J_{3}) = \left (
      \begin{matrix}
    \cos\theta & -\sin\theta & 0 \\
    \sin\theta & \cos\theta & 0 \\
    0 & 0 & 1
      \end{matrix}
      \right),
\end{equation}
where $J$ is the generator of total angular momentum, and $\theta=\arctan(\frac{x_2}{x_1})$.
With the help of the $SO(3)$ analogue of Eq.(\ref{omegai}), the superfluid velocities Eq.(\ref{omegab})are
readily calculated to be
\begin{eqnarray} \label{omegaB}
 \omega^{3}_{1}  &=& -(\partial_{1}R_{1k})R_{2k} = \frac{x_{2}}{r^{2}} = \frac{2\mu}{\hbar c^2}E_2 \nonumber \\
\omega^{3}_{2}  &=& -(\partial_{2}R_{1k})R_{2k} =-\frac{x_{1}}{r^{2}} = -\frac{2\mu}{\hbar c^2}E_1 \nonumber \\
\omega^{1}_{3} &=& -(\partial_{3}R_{2k})R_{3k} = 0 \nonumber \\
\omega^{2}_{1} &=& -(\partial_{3}R_{1k})R_{3k} = 0,
\end{eqnarray}
where $r^{2} = x_{1}^{2} + x_{2}^{2}$. Since the groups $SO(3)$ and $SU(2)$ give the same
equations of motion Eq.(\ref{eomgeneral}), we see that the Ansatz Eq.(\ref{u1R}) satisfies these as
well, giving a spin-Hall current for the $z$-polarised spin. In other words, in \he -B is a
possible candidate for our quantised spin vortex.

This result can also be understood by topological means, in the following way. The equation of
motion for the $SU(2)$ case tells us, that the vacuum manifold for the spin becomes $U(1)$
instead of $SO(3)\simeq SU(2)$. Only if we were allowed to change the orientation of the wire,
described by a point on $S^2$, we would obtain the full $SO(3)$. This is the translation of
the mathematical fact that $SO(3)/S^2 \simeq U(1)$, merely saying that a rotation is fixed by
an axis of rotation and the angle of rotation about that particular axis. The implication is
that we need to calculate the fundamental group of $G_B/S^2$ instead of $G_B$ itself:
\begin{equation}
\pi_1(SO(3)/S^2) = \pi_1(U(1)) = \mathbb{Z}\, ,
\end{equation}
leading to the existence of vortices in a cylindrical set up, i.e., the inclusion of radial
electric fields induces vortices.

There is however one effect which destroys our spin vortex solution. This effect, known as
dipolar locking, will be discussed in the next section.

\subsection{Dipolar locking} \label{sect:diplock}

In the 1970s, Leggett described in his seminal article about \he\  many important properties of
this interesting system \cite{leggetthe}. One of them is how the spin part of the condensate
wave function $\Psi(\vec x)$ interacts with its orbital motion by a $\vec S \cdot \vec L$
interaction. According to Leggett, the contribution of the Cooper pairs to the dipolar energy
is
\begin{align}
E_{dip} &=& -g_{dip} \int d\vec x  \frac{1}{x^3}\left(  |\Psi(\vec x)|^2  - 3 |\vec x \cdot \Psi(\vec x)
|^{2} \right) \nonumber \\
  &=& g_{dip} \int \frac{d\Omega}{4\pi} 3 | \hat{n}\cdot (A_{\alpha i}
n_{\alpha}) |^2 - \mbox{constant},
\end{align}
remembering that the spin order parameters carry a spatial index, cf. Eq.'s(\ref{AA}), (\ref{AB}).
We used the notation $\hat{n} = \frac{\vec x}{|x|}$. On inserting the order parameters Eq.'s(\ref{AA})
and (\ref{AB}), we obtain for both phases the dipole locking Lagrangians
\begin{align}
L_{dip, B} &=& -g_{dip} \left( (\mbox{Tr} R)^2 + \mbox{Tr}(R)^2 \right), \nonumber \\
L_{dip, A} &=& -g_{dip} (\hat{l}\cdot\hat{d})^2.
\end{align}

For the \he-A part, we do not need to solve the equations of motion in order to infer that the
orbital and spin vector wish to be aligned. For the B-phase, we give a derivation of the
Leggett angle. A general matrix $R\in SO(3)$ can be described by three Euler angles. For the
trace, only one of them is important, let's say it is called $\theta$. Then
\begin{equation}
L_{dip, B} = -g_{dip} \left\{ (1+ 2 \cos\theta  )^2 + 2(\cos^2\theta -\sin^2\theta )\right\},
\end{equation}
which leads to the static equation of motion
\begin{equation}
0 = \frac{d L_{dip, B}}{d \theta} = 4 \cos \theta -1 ,
\end{equation}
with the Leggett angle as solution,
\begin{equation}
\theta_L = \arccos(-\frac{1}{4}) \simeq 104^o.
\end{equation}

The Leggett angle tells us that one degree
of freedom is removed from the order parameter of $^3$He-B
so that
\begin{equation}
SO(3)_{rel} \rightarrow G_{B, dip}=S^2,
\end{equation}
but $\pi_{1}(S^2)=0$, as any closed path on the sphere can be continuously shrunk to a point.

\begin{figure}[ht!]
      \centering
      \rotatebox{0}{
    \resizebox{5.3cm}{!}{%
      \includegraphics*{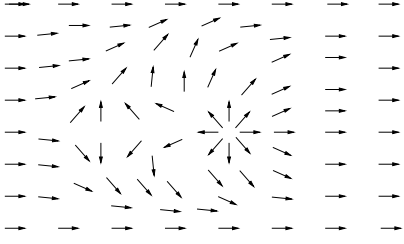}}}
      \caption{ The destruction of the spin vortex by dipolar locking. The 
               $U(1)$ degree of freedom is indicated by an arrow. In the center where the 
               electric field is located, 
               the angle follows a vortex configuration of unitwinding number, corresponding
               to one charge quantum. 
               Since the electric field, 
               decaying as $\frac{1}{r}$, is not able to compete with the dipolar locking 
               at long distances, the $U(1)$ angle becomes fixed at the Leggett
               angle, indicated by a horizontal arrow. }
              \label{vortexdestruction}
    \end{figure}

Now we can also understand that dipolar locking destroys vortices, even in a cylindrical set
up, i.e. with a radial electric field, since
\begin{equation}
\pi_1(G_{B, dip}/S^2) = \pi_1(e) = 0\, .
\end{equation}
The ``division'' by the manifold $S^2$ translates the fact that different vortices in the
\he-B manifold are only equivalent to each other up to different orientations of the
cylindrical wire, being described by $S^2$. Another way to understand the destruction of
vortices  beyond the dipolar length, is that the $U(1)$ vortex angle $\theta$ is fixed to the
Leggett angle, as depicted in figure \ref{vortexdestruction}.

The fact that the vortices are destroyed, even though the spin-orbit coupling energy is higher
than the dipolar locking energy \cite{minvol}, is due to the fact that small energy scales do
play a role at large distances. This is similar to spontaneous symmetry breaking in, for
example, an XY-antiferromagnet. A small external field is enough to 
stabilize domain walls
at long wavelengths.

\subsection{\he-A} \label{hea}

In the discussion of the pure spin superfluids and of \he-B, we used the fact that the order
parameter has a matrix structure, namely $SU(2)$ and $SO(3)$, respectively. For the $SU(2)$
case we had to transform from the fundamental spinor representation to the adjoint matrix
representation. Since both representations are $SU(2)$, the physics did not change
fundamentally. The resulting equations of motion were equations for group elements $g$, with
the ramification that spin vortex states lower the energy with respect to the trivial
solution, cf. Eq.(\ref{spinvortexlowersenergy}). As a result, the vacuum manifolds in both
cases become $U(1)$ instead of $SU(2)$ (pure spin superfluid) or $SO(3)$ (\he -B without
dipolar locking). The topological protection of the spin vortex solution followed from the
fact that $U(1)$ is characterised by the winding numbers, $\pi_1(U(1))=\mathbb{Z}$.

For the case of \he-A, matters are different, since the spin order parameter for \he-A is a
vector in $S^2$ instead of a matrix in $SO(3)$. Although $SO(3)$ acts on $S^2$, these
manifolds are not the same. What we will prove is that as a result, spin vortices do
\emph{not} lower the energy in the presence of an electric field, as opposed to the \he-B and
pure spin superfluids. The consequence is that the vacuum manifold remains $S^2$, and since
$\pi_1(S^2)=0$, spin vortices are not protected. The presence of dipolar locking will not
change matters.

Let us prove our assertions by deriving the equations of motion from the Lagrangian for \he-A.
The free energy functional \cite{volovikexo} for $^3$He-A is quite analogous to that of a liquid
crystal\cite{degennes}, as the A phase is both a superfluid and a liquid crystal in some sense.
Besides the bulk superfluid energy, there are also gradient energies present in the free
energy, of which the admissible terms are dictated by symmetry:
\begin{align}
\begin{aligned}\label{fgrad}
F_{grad} = \gamma_1 (\partial_i A_{\alpha j}) (\partial_i A_{\alpha j})^* + & \gamma_2
(\partial_i A_{\alpha
i})(\partial_j A_{\alpha j})^* \\
+ \gamma_3 (\partial_i A_{\alpha j})&(\partial_j A_{\alpha i})^*\\
A_{\alpha i} = \Delta_A \hat{d}_\alpha e^{i\phi_{rel}}&( \hat{e}^{(1)}_{i} + i
\hat{e}^{(2)}_{i}) \,.
\end{aligned}
\end{align}
 This then leads to
\begin{align} \label{FgradA}
F^{London}_{grad} &=&
 \frac{1}{2}K_{ijmn}
\partial_{i}\hat{e}_{m}\partial_{j}\hat{e}_{n}
+ C_{ij}(v_{s})_{i}\epsilon_{jkl}\partial_{k}\hat{e}_{l}
\nonumber \\
 & & +\frac{1}{2}\rho_{ij}
(\partial_{i}\hat{d}_{\alpha})(\partial_{j}\hat{d}_{\alpha})
 +g_{dip}(\hat{d}_{\alpha}\hat{e}_{\alpha})^{2}.
\end{align}
The coefficients $K_{ijmn}$ and $C_{ij}$ are the liquid crystal like
 parameters\cite{degennes}.The superfluid velocity $v_s$ is the Abelian superfluid
velocity coming from the relative $U(1)$ phase.

We are going to prove that \he-A does not have topologically stable spin vortices, and that
dipolar locking does not stabilize these. Generically, the spin stiffness tensor $\rho_{ij}$
is given by \cite{volovikexo}
\begin{equation}
\rho_{ij} = \rho^{||}\hat{l}_{i}\hat{l}_j + \rho^{\perp}\left(\delta_{ij} -
\hat{l}_{i}\hat{l}_j\right),
\end{equation}
but it becomes fully diagonal when we neglect anisotropies in the spin wave velocities, i.e.,
$\rho^{||}=\rho^{\perp}$. We also assume that the $K_{ij,mn}$ and and $C_{ij}$ are fully
diagonal, since this will not change the nature of the universal low energy physics. Including
now spin-orbit coupling and kinetic terms the $^3$He-A Lagrangian is
\begin{align}
        \begin{aligned}
          \label{lagheA2}
          L^A(\psi_{\alpha j},\vec E,\vec B) = -\frac{\hbar^2}{2mc^2}
          \left\{ |\partial_0 \hat e_j |^2
          + \left(\partial_0
          d_\alpha\right)^2 + \right. \\
    \left. \frac{2\mu m n_s}{\hbar^3 c}
          \epsilon_{\alpha\beta\gamma}\hat d_\beta\partial_0\hat
          d_\gamma B_\alpha \right\} \\
          + \frac{\hbar^2}{2m} \left\{ |\partial_i \hat
          e_j|^2  + \left(\partial_i d_\alpha\right)^2 -
          \frac{2\mu m n_s}{\hbar
          c^2}\epsilon_{\alpha\beta\gamma}\epsilon_{\alpha ik}\hat
          d_\beta\partial_i\hat d_\gamma E_k\right\} +\\
          \frac{1}{8\pi}\left(E^2 - B^2\right) - \frac{1}{2} g_{dip}
          \left( \hat d \cdot \hat l\right)^2.
        \end{aligned}
      \end{align}

The strategy for solving the equations of motion is as follows: first we demonstrate that a
spin vortex is possible without dipolar locking, but that it does not gain energy with respect
to the constant solution. Then we show that the spin vortex is not stabilized by switching on
the dipolar locking.

Without dipolar locking a spin-only action is obtained, leading to an equation of motion which
resembles Eq.(\ref{eomgeneral}),
\begin{equation}\label{deq}
\partial_i \left[ \partial_i d_j -  \frac{2\mu m n_s}{\hbar
          c^2}\epsilon_{\alpha ik} (\epsilon_{\alpha\beta j})d_\beta E_k \right] = 0.
\end{equation}
Let us choose a reference vector $D_\nu$, such that $d_j = R_{j\nu} D_\nu$. Again, $R$ is an
$SO(3)$ matrix, describing the superfluid phase of the $S^2$ variable $d$. In this way, the
equation of motion for the group element $R$ reads
\begin{equation}\label{RAeq}
\partial_i \left[ \partial_i R_{j\nu} -  \frac{2\mu m n_s}{\hbar
          c^2}\epsilon_{\alpha ik} (\epsilon_{\alpha\beta j}) R_{\beta\nu}  E_k \right] = 0.
\end{equation}
Using cylindrical coordinates, the demonstration that the spin vortex Ansatz for $R$ is a
solution to this equation of motion is analogous to the proof that a spin vortex exists in \he
-B, cf. Eq.(\ref{omegaB}). On the other hand, this equation also admits a constant $R$, i.e., 
Eq.(\ref{deq}) admits a constant $D_\mu$ as well. Substituting both solutions back into the
energy functional Eq.(\ref{lagheA2}), no energy differences between the
spin vortex and the constant solution show up. In mathematical terms, the vacuum manifold in the
presence of a cylindrical electric field remains $S^2$. In plain physics language: the
electric field does not prevent phase slips to occur.

The presence of dipolar locking makes matters even worse, since the equations of motion become
equations of motion for $e$ and $d$ involving dipolar locking,
\begin{eqnarray}\label{eomgeneraldiplock}
& & \frac{\hbar^2}{2m}\partial_i^2 \hat{e}^{(1)}_j =
- g_{dip} (\epsilon_{abc}\hat{e}^{(1)}\hat{e}^{(2)}_c \hat{d}_a ) \epsilon_{kjm}\hat{e}^{(2)}_m \hat{d}_\alpha \nonumber  \\
& & \frac{\hbar^2}{2m}\partial_i^2 \hat{e}^{(2)}_j =
- g_{dip} (\epsilon_{abc}\hat{e}^{(1)}\hat{e}^{(2)}_c \hat{d}_a ) \epsilon_{kmj}\hat{e}^{(1)}_m \hat{d}_\alpha \nonumber  \\
& &  \partial_i \left[ \partial_i d_j -  \frac{2\mu m n_s}{\hbar
          c^2}\epsilon_{\alpha ik} (\epsilon_{\alpha\beta j})d_\beta E_k \right] \nonumber \\
& &          = - 2 g_{dip} (\epsilon_{abc}\hat{e}^{(1)}\hat{e}^{(2)}_c \hat{d}_a ) \epsilon_{jlm}\hat{e}^{(1)}_l \hat{e}^{(2)}_m. \nonumber \\
 \end{eqnarray}
It is clear that in general, a vortex configuration for $\hat{d}$ is not a solution, since the
left hand side of the equation for $\hat{d}$ is annihilated, whereas the right hand side is
not. Instead, the orbital and spin vectors will perform some complicated dance, set in motion
by the electric field.

The verdict:  our charge trapping experiment  will not work employing \he-A.

\subsection{Baked Alaska}

In the search for an experimental realisation of the proposed charge trapping experiment, it
turned out that \he-B admits spin vortex solutions only at short wavelengths. But if there
were a way to circumvent dipolar locking in some ideal world, nothing would stop us from
performing the actual experiment.

Or... does it? It turns out that the numbers which Nature gave us, conspire to obstruct
matters. It is really hidden in the fact that electric fields are so strong, and spin-orbit
coupling so weak. Let us first confess that in the previous considerations, we did not regard
a very important part of our charge trapping device, namely, the wire itself. The charge
stored on it is hugely repelling indeed, giving rise to an enormous charging energy.

First, we calculate the Coulomb energy stored in the wire. Let $\rho(x)$ be the charge density
distribution, which we approximate by a step function of the radius. Then,
\begin{eqnarray}
W_{\mbox{\scriptsize Coulomb}} &=& \frac{1}{8\pi\epsilon_{0}} \int \frac{\rho(x)\rho(x)}{\|
\mathbf{x}-\mathbf{x}'\|} d\mathbf{x} d\mathbf{x}' \nonumber \\
&=&\frac{1}{8\pi\epsilon_{0}} \frac{Q^{2}_{\mbox{\scriptsize tot}}}{\pi a^{2}L} I.
\end{eqnarray}
We integrated over the center-of-mass coordinate, and (with the definitions $\mathbf{u}
=\mathbf{x}-\mathbf{x}'$ and $r=L/a$) we introduced
\begin{eqnarray}
I &\equiv& \int_{0}^{L}du_{z} \int_{0}^{a}2\pi du_{\perp}u_{\perp} \frac{1}{\|\mathbf{u}\|}
\nonumber \\
&=& 2\pi \left\{ -\frac{1}{2}L^{2} +
a\int_{0}^{L}du_{z}\sqrt{1+\left(\frac{u_{z}}{a}\right)^{2}} \right\} \nonumber \\
&=& 2\pi \left\{ -\frac{1}{2}L^{2}  +\frac{a^{2}}{2}\left( q\sqrt{1+q^{2}} + \ln
(q+\sqrt{1+q^{2}})
\right)\right\} \nonumber \\
&\simeq & 2\pi \frac{a^2}{2} \ln(2q) \ \ \  \  \mbox{for $L>>a$}.
\end{eqnarray}
We used the standard integral $\int d\tau \sqrt{1+\tau^2} = \frac{1}{2}\tau\sqrt{1+\tau^2} +
\frac{1}{2}\ln (\tau + \sqrt{1+\tau^2})$. Hence
\begin{equation}
W_{\mbox{\scriptsize Coulomb}} = \frac{1}{8\pi\epsilon_{0}} \lambda^{2} L \ln\left(
\frac{2L}{a}\right).
\end{equation}
For the parameters under estimation, $W_{\mbox{\scriptsize Coulomb}}/L \simeq  1 $J/m, which
is really enormous, since the coupling constant of electric fields is so huge.

The question is now if the superfluid is strong enough to keep the charge trapped. Indeed, if
it doesn't, the system can lower its energy by simply discharging the wire, causing a big
spark, and destroying the superfluid. This is analogous to magnetic flux trapping in
superconducting rings with the Aharonov-Bohm effect \cite{aharonovbohm}. The flux trapped in a
ring is a metastable state, but the superconducting condensate is strong enough to keep it
there.

However, spin-orbit coupling is too weak to do so with our Aharonov-Casher analogue. In fact,
the only thing the system needs to do, is to destroy the spin superfluid, not in the whole
container, but just a small strip of the order of the coherence length $\xi$, which is of the
order of $0.01 \mu m$ \cite{seppala}.
\begin{figure}[ht!]
      \centering
      \rotatebox{0}{
    \resizebox{5.3cm}{!}{%
      \includegraphics*{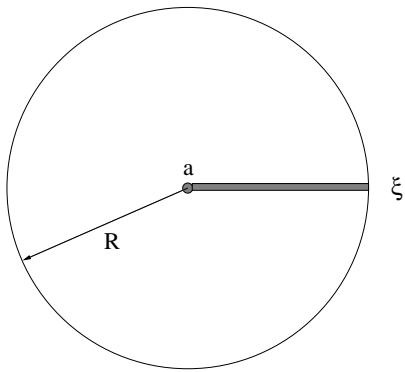}}}
      \caption{ View from the top of our container. The container radius is $R$, and the wire
       has radius $a$. Now, the Coulomb energy of the wire has to make a tiny region of superfluid
       normal again, in order to make phase slips happen, removing the topological constraint. The
       region in which this should happen, needs to be of the width of the coherence length $\xi$,
       but it has to extend over the whole radius of the container.}
      \label{phaseslip}
    \end{figure}
We now need to estimate the energy density of the fluid. To do this, we perform Landau theory
for the superfluid order parameter $\psi$,
\begin{equation}
\delta F = \int \left\{  a|\psi|^2 + \frac{1}{2}b|\psi|^{4} \right\} d\mathbf{x}.
\end{equation}
This expression is zero when there is no superfluid. There is no kinetic term, since $\psi$ is
parallel transported by the electric field: indeed, if it satisfies the equations of motion,
the kinetic term vanishes, cf. Eq. (\ref{eomgeneral}). Hence, we are only left with the
potential energy terms. From Landau theory, we know the saddle point value for $\psi$ in terms
of $a=\alpha(T-T_{ c})$ and $b$, viz.,
\begin{equation}
|\psi|^{2} = \frac{-a}{b} \Rightarrow \delta F = -V \frac{\alpha^{2}}{b}(T-T_{ c}),
\end{equation}
where $V=\pi R^{2}L$ is the volume of the container. Note that $R$ is the unknown variable in
our problem. From Landau and Lifschitz we obtain the BCS-parameters
\begin{equation}
a(T) = \frac{6\pi^2}{7\zeta(3)}\frac{k_{ B}T_{ c}}{\mu}(k_{B}T_{c}) \left(1 -
\frac{T}{T_{c}}\right),\mbox{ \ \ \  } b = \alpha\frac{k_{B}T_{ c}}{\rho},
\end{equation}
where $\rho$ is the superfluid density. For low temperatures $T<<T_{ c}$ we have $\mu \simeq
\varepsilon_{F}$, s
\begin{equation}
\delta F \simeq 3.52 (nk_{B}T_{ c}V)\frac{k_{ B}T_{ c}}{\varepsilon_{ F}}.
\end{equation}
We use experimental values \cite{seligman} $\varepsilon_{ F}/k_{ B} = 0.312 K$ and
$T_{c}=3 mK$. From the Fermi gas relation $\rho = p^{3}_{ F}/3\pi^2\hbar^2$ we then obtain
$\rho\approx 15$ mol/liter. This leaves us with an estimate
\begin{displaymath}
\frac{\delta F}{V}  \sim 34 \ \mbox{J}/\mbox{m}^3.
\end{displaymath}
The question we need to ask is: how big does the container radius $R$ need to be, in order to
remain in the metastable, charge trapped state? Per length unit $L$, the estimate is
\begin{equation}
\frac{W_{Coulomb}}{L}  = \frac{\delta F}{V} R \xi.
\end{equation}
Due to the enormously small $\xi$ and the enormously big $W_{Coulomb}$, this leads to a truly
disappointing radius of
\begin{equation}
R \simeq 1000 km,
\end{equation}
enough to cover Alaska, and much more than the total amount of He \ on Earth (180 liters).
There might be enough He \ on the Moon, but still it is a ``only  in your wildest dreams''
experiment. Is there no way out? In the concluding section, we give a direction which might
provide some hope.

\section{Outlook: organic superconductors}

In the previous section, we have seen that the small spin-orbit coupling energy and the big
electric fields are disastrous. This is due to the fact that the coherence length $\xi$ is
small. In turn, the reason for that is that in Landau theory, $\xi \propto
\frac{1}{\sqrt{m}}$. In other words, the heavier the constituent particles, the worse things
get. So we need to look for lighter things. The first candidate would be electrons, since they
are 5000 times lighter. However, as they are charged, charge effects highly overwhelm the
whimpy spin-orbit coupling effects. So we need something made out of electrons, having however a huge
gap for charge excitations: we need a spin superfluid made out of a Mott insulator. Does this exist?

\begin{figure}[ht!] 
      \centering
      \rotatebox{0}{
    \resizebox{8.3cm}{!}{%
      \includegraphics*{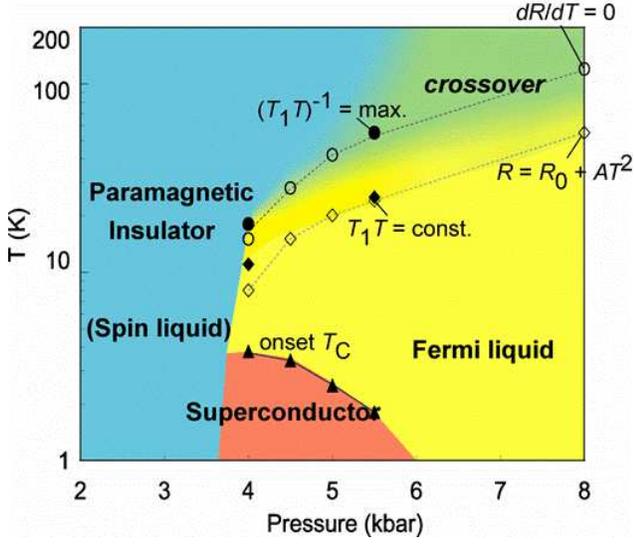}}}
      \caption{ The phase diagram of the highly frustrated $\kappa$-(ET)$_{2}$Cu$_{2}$(CN)$_{3}$,
               as proposed by Kanoda \cite{kanoda05}. The spin liquid state shows linear specific heat,
               which might signal the presence of a spinon Fermi surface. This would amount to making
               a spinon Fermi liquid out of an insulator. Then the interesting possibility
                is that this spinon metal might
               be unstable against an $S=1$ spin superfluid.}
             \label{kanoda}
    \end{figure}

In recent years, there have been many advances in the research on highly frustrated systems on
triangular lattices \cite{moessnersondhi}, which are realised in organic compounds. In the
last two years,  Kanoda \textit{et al.} have done specific heat measurements in the spin
liquid phase of the organic superconductor $\kappa$-(ET)$_{2}$Cu$_{2}$(CN)$_{3}$, see Figure
\ref{kanoda}. Although the spin liquid state is known to be a featureless paramagnet, the
specific heat shows a linear behaviour as a function of temperature \cite{kanoda03, kanoda05}.

The linear behaviour has led theorist P.A. Lee to the idea that this might be caused by
fermionic spinons forming a Fermi surface \cite{leeorganicSL}. It is plausible that at low
energy scales, a BCS-like instability of the Fermi surface might give rise to an $S=1$ spinon
condensate. This would then be the desired spin superfluid made out of a Mott insulator.
 The theoretical complication is that due to the
$SU(2)$ slave theories developed by Lee and Wen \cite{lnw}, there will be transversal gauge
degrees of freedom, blocking the triplet channel, which should give rise to some scepticism
about whether the organics are able to become a triplet superfluid.
Whether or not this is the case, to our opinion, the idea of charge trapping provides
a good motivation to pursue the
BCS-instability towards a triplet state of the spinon metal further.

\appendix\section{Useful formulas}

In order to obtain Eq's(\ref{cons0}) and (\ref{joe}) we
calculate the expressions
\begin{align}
&D_0 \psi =  \left[ \frac{\partial_0\rho}{2\rho} + i \left( \partial_0
\theta -e A_0 + u_0^aS^a -q A^a_0 S^a\right)\right]\psi\\
&\vec D \psi = \left[ \frac{\vec \nabla \rho}{2 \rho} + i \frac{m}{\hbar}
\left( \frac{\hbar}{m}\vec \nabla \theta -\frac{e \hbar}{m} \vec A
+ \vec u^aS^a -\frac{q\hbar}{m} \vec A^a S^a\right)\right]\psi
\end{align}
We also obtain
\begin{align}\nonumber
& \vec D^2 \psi =  \\ \nonumber
& \psi \left\{ \frac{\vec \nabla^2 \rho}{2\rho} - \frac{1}{4}
\left( \frac{\vec \nabla \rho}{\rho}\right)^2 -\frac{m^2}{\hbar^2}
\left(\vec u -\frac{e \hbar}{m} \vec A+ \vec u^a S^a - \frac{q\hbar}{m}
\vec A^a S^a\right)^2\right\}\\
& + \psi \left\{ i\frac{m}{\hbar \rho}\vec \nabla \cdot \left[ \rho \left(
\vec u -\frac{e \hbar}{m} \vec A + \vec u^a S^a - \frac{q\hbar}{m}
\vec A^a S^a
\right)\right] \right\}
\end{align}
and substitute them in the Pauli equation Eq.(\ref{paulilike}). In order
to obtain Eq.(\ref{consna}), we multiply the Pauli-like equation
by $\tau^a/2$ and use the expressions
\beq
\frac{\tau^a}{2}\psi = \psi S^a
\eneq
\begin{widetext}
\begin{align}
&\frac{\tau^a}{2} D_0\psi = \psi\left\{ \frac{\partial_0\rho}{2\rho}S^a
+ i(\partial_0\theta  -e A_0)S^a + \frac{i}{4}u_0^a - \frac{i}{4}qA_0^a -
\frac{1}{2}\epsilon^{abc}u_0^bS^c + \frac{1}{2}
q\epsilon^{abc}A_0^bS^c \right\} \\
&\nonumber \frac{\tau^a}{2} \vec D^2\psi = \psi\left\{
\left[\frac{\nabla^2\rho}{2\rho} -
\frac{1}{4}\left(\frac{\nabla\rho}{\rho}\right)^2\right] S^a -
\frac{m^2}{\hbar^2}\left[ \vec u^2 + \frac{1}{4}\left( \vec u^b -
\frac{\hbar q}{m}\vec A^b\right)^2\right] S^a -
\frac{im^2}{\hbar^2}\epsilon^{abc}\vec u\cdot\left(\vec u^a -
\frac{\hbar q}{m}\vec A^a\right) S^c \right\}\\
&\nonumber \qquad\quad\; + \psi \left\{ -\frac{m^2}{2\hbar^2} \vec
u\cdot\left( \vec u^a - \frac{\hbar q}{m}\vec A^a \right) +
\frac{im}{\hbar\rho}\nabla\cdot\left( \rho\vec u \right) S^a +
\frac{im}{\hbar\rho}\nabla\cdot\left[ \rho\left(\vec u^a - \frac{\hbar
q}{m}\vec A^a\right) \right] \right\} \\
& \qquad\quad\; + \psi \left\{ -
\frac{m}{2\hbar\rho}\epsilon^{abc}\nabla\cdot\left[ \rho\left( \vec
u^b - \frac{\hbar q}{m}\vec A^b \right) \right] S^c -
\frac{imq}{4\hbar}\epsilon^{abc}\vec u^b\cdot\vec A^c +
\frac{mq}{2\hbar}\left( \vec u^b\cdot\vec A^b - \vec u^a\cdot\vec
A^b\right) S^b \right\} \, .
\end{align}
\end{widetext}

\noindent \textbf{ Acknowledgements:}  We acknowledge helpful
discussions with M. Mostovoy, R. Jackiw, S.C. Zhang, A.V. Balatsky and N. Nagaosa.
This work was financially supported
by the Dutch Science Foundation NWO/FOM.

\end{document}